\documentclass[aps,prd,onecolumn,11pt,groupedaddress,nofootinbib,showpacs]{revtex4}
\usepackage{amsmath}
\usepackage{pstricks,pst-node}
\usepackage{supertabular}

\begin{document}
\title{ Replica symmetric spin glass field theory
}

\author{T.~Temesv\'ari}
\email{temtam@helios.elte.hu}
\affiliation{
Research Group for Theoretical Physics of the Hungarian Academy of Sciences,
E\"otv\"os University, P\'azm\'any P\'eter s\'et\'any 1/A,
H-1117 Budapest, Hungary}

\date{\today}

\begin{abstract}
A new powerful method to test the stability of the replica symmetric
spin glass phase is proposed by introducing a replicon generator function
$g(v)$. Exact symmetry arguments are used to prove that its extremum is
proportional to the inverse spin glass susceptibility. By the idea of
independent droplet excitations a scaling form for $g(v)$ can be derived,
whereas it can be exactly computed in the mean field Sherrington-Kirkpatrick
model. It is shown by a first order perturbative treatment that the replica
symmetric phase is unstable down to dimensions $d\lesssim 6$, and the mean
field scaling function proves to be very robust. Although replica
symmetry breaking is escalating for decreasing dimensionality, a mechanism
caused by the infrared divergent replicon propagator may destroy the mean
field picture at some low enough dimension.
\end{abstract}


\maketitle

\section{Introduction}

The theory of spin glasses has by now become more than three decades old,%
\footnote{For reviews at different stages of the story, see
\cite{BinderYoung,MePaVi,FischerHertz,Nishimori} and the collection of
papers in \cite{SG98} about different aspects of spin glass theory.}
even if one consideres the archetypal model by Edwards and Anderson
\cite{EA} as its beginning. Nowadays spin glasses are regarded as the
prototypes of complex systems showing highly nontrivial equilibrium
properties (massless glassy phase without long-range order, nonergodic
Gibbs state with a complicated free energy surface, lack of self-averaging
of some quantities, etc. \ldots) on the one hand, and a very slow aging
dynamics with diverging relaxation times on the other hand.
The state of affairs in the understanding of basic issues of the equilibrium
glassy phase is still controversial, as two rather different theories have
emerged for the low-temperature phase of the short-range Ising spin glass:
One of these two schools extends the idea of replica symmetry breaking (RSB),
invented by Parisi (see \cite{MePaVi}, and references therein)
for the mean field version of the model --- i.e.\ for the
Sherrington-Kirkpatrick (SK)
model \cite{SK} ---, to the short-range case in the physical dimensions. The other one,
the so called scaling or droplet picture \cite{McMillan,FiHu86,BrMo86,FiHu88},
claims --- at least in its most radical version --- that RSB is restricted
to the SK model which, with its infinite coordination number, can be relevant
only in infinite dimension ($d=\infty$), or for long-range models in finite
$d$. Before displaying a list of the most important conflicting statements,
we present the model and give some definitions.

The Hamiltonian of the Edwards-Anderson (EA) model on a $d$-dimensional
hypercubic lattice with the nearest neighbour quenched interactions
$J_{ij}$'s taken independently from a symmetric Gaussian distribution is
\cite{EA}
\[
\mathcal{H}_{{J}}=-\sum_{(ij)}J_{ij}\,S_iS_j-H\sum_{i}S_i,
\]
with $N$ Ising spins sitting on the lattice sites, and $N\to \infty$ in the
thermodynamic limit.
(The second term contains the external magnetic field $H$,
although it is zero throughout this paper.)
The distribution $P_J(q)$ of the site overlaps
$q_{SS'}=\frac{1}{N}\sum_iS_iS'_j$, while picking out the spin configurations
$S$ and $S'$ independently by the Gibbs-measures with the same
$\mathcal{H}_{{J}}$'s, is usually considered as the order parameter
{\em function\/}. The maximum value of the $P_J(q)$'s support is the
EA order parameter $q_{\text{EA}}$, although originally
$q_J\equiv\frac{1}{N}\sum_i\langle S_i\rangle ^2=\int dq\,q\,P_J(q)$ was proposed as
$q_{\text{EA}}$ \cite{EA}, which generally depends on the realization $J$.
Here and in subsequent definitions $\langle \dots \rangle$ means thermal
average by the Gibbs measure $\sim e^{-\mathcal{H}_{{J}}/kT}$, whereas
$\overline{\rule{0pt}{6pt}\dots}$
stands for averaging over the quenched disorder. We
have $P(q)\equiv\overline{P_J(q)}$.
The following three 2-site 
(zero momentum) correlation functions are useful to define:
\begin{equation}\label{G's}
\begin{aligned}
G_1&=\frac{1}{N}\sum_{ij}\overline{\langle S_iS_j\rangle^2}-N\,
\overline{q_J}^2\\
G_2&=\frac{1}{N}\sum_{ij}\overline{\langle S_iS_j\rangle
\langle S_i\rangle \langle S_j\rangle }-N\,\overline{q_J}^2\\
G_3&=\frac{1}{N}\sum_{ij}\overline{\langle S_i\rangle ^2
\langle S_j\rangle^2}-N\,\overline{q_J}^2.
\end{aligned}
\end{equation}
Two linear combinations of these are especially important to characterize
the spin glass phase \cite{BrMo86}:
\begin{align}
G_{\text{R}}&=G_1-2G_2+G_3=\frac{1}{N}\sum_{ij}\,
\overline{(\langle S_iS_j\rangle-\langle S_i\rangle\langle S_j\rangle)^2}
\,\equiv\, \chi_{\text{EA}}\label{chiEA}\\
\intertext{and}
G_{\text{L}}&=G_1-4G_2+3G_3.\notag
\end{align}
The subscripts R and L are for ``replicon'' and ``longitudinal'', and
$G_{\text{R}}=\chi_{\text{EA}}$ is the famous Edwards-Anderson (or
spin glass) susceptibility whose inverse is the so called replicon
mass $2m_1=\chi_{\text{EA}}^{-1}$,
which is used throughout this paper to test the
stability of the spin glass phase.

In the following list, it is attempted to review the most important conflicting
statements provided by the two rival theories:
\begin{enumerate}
\item
The Gibbs measure is decomposed into an infinite number of pure states, not
related by any symmetry, in RSB. The droplet picture is like the ferromagnet:
there is one pair of pure states related by spin inversion. As a result, a
trivial probability distribution of the
overlaps between equilibrium spin configurations arises,
with a pair of delta functions at
$\pm q_{\text{EA}}$.
A nontrivial and non-self-averaging
[i.e.\ $P_J(q)\not=P(q)$] overlap 
distribution follows
from the Parisi theory \cite{MePaVi}, with a continuous part for a range of
overlaps smaller than $q_{\text{EA}}$ plus the delta function for
self-overlap. $q_{\text{EA}}$ is, however, self-averaging in both theories.
\item
The ergodic components, or pure states, in the Parisi theory are
ultrametrically organized \cite{MePaVi}.
\item
The average system-size low-energy excitations, when modelling in a
finite size system, are proportional to $N^{\Theta/d}$,
$\Theta>0$ being the stiffness exponent, and the surface of
the compact cluster of flipped spins,
i.e.\ of the ``droplet'', is a fractal with $d_s<d$ in the
droplet picture \cite{FiHu86,FiHu88}. On the other hand, the same type of
exitations are thought to be of order unity, and they are space filling
($d_s=d$) in RSB. 
\item
The external magnetic field $H$ destroys the spin glass transition in the
droplet picture, whereas the spin glass phase is bounded by the
Almeida-Thouless (AT) line $H_{\text{AT}}(T)$ in RSB
\cite{AT}. Along the AT line
$\chi_{\text{EA}}$ of (\ref{chiEA}) is infinite, that is the replicon mass
is zero and starts to become negative, signalling the end of the replica
symmetric (RS) paramagnetic phase, and the onset of RSB.
\item
There is one feature both theories share, namely the spin glass susceptibility
$\chi_{\text{EA}}$ is infinite,
in the thermodynamic limit, in the {\em whole\/} glassy phase, i.e.\
it is a marginally stable massless phase. The details, however, already
differ, as the divergence of $\chi_{\text{EA}}$ with $N$ is
\begin{equation}\label{point 5}
\chi_{\text{EA}}\sim
\begin{cases}
N & \text{for RSB,}\\
N^{1-\Theta/d} & \text{droplet picture, see \cite{BrMo86,FiHu88}.}
\end{cases}
\end{equation}
\end{enumerate}

The discussion of the above points may be started by considering
``replica equivalence,'' a set of identities between joint distributions of
the site (or standard) overlaps, and rigorously proved for the SK model
\cite{G,AiCo98,GG}.
Its first derivation was based on the ultrametric property of
the Parisi solution of the SK model \cite{Mezard_et_al}, it is, however,
more generic than ultrametricity \cite{Parisi04}.
Applying replica equivalence to the susceptibilities in (\ref{G's}),
we can easily derive for the leading $O(N)$ term [see also (\ref{chiEA})]:
\[
G_2=\frac{1}{2}G_1,\qquad G_3=\frac{1}{3}G_1\qquad \text{and}\qquad
\chi_{\text{EA}}=\frac{1}{3}G_1=N\,\frac{1}{3}\bigg[\int dq\,q^2 P(q)-
\Big(\int dq\, qP(q)\Big)^2\bigg].
\]
This formula is in complete accordance with point 5 above for both scenarios.
It is quite interesting, and not completely understood, that the ratio
$1:\frac{1}{2}:\frac{1}{3}$ for the leading terms of the $G$'s,
which are $O(N^{1-\Theta/d})$, in the droplet case remains true, although
not because of replica equivalence.

Replica equivalence has recently been rigorously proven for a large class of
models, including the EA model, for the {\em link\/} overlap
$q^l_{SS'}=\frac{1}{d\,N}\sum_{(ij)}S_iS_j\,S'_iS'_j$
\cite{Contucci03,CoGi05,CoGi06}. A very recent numerical study
\cite{CoGiGiVe06} for the 3-dimensional model provided strong support
that the two overlaps (site and link) are equivalent in the description
of the spin glass phase of the EA model. Considering also the fact that
replica equivalence is a very plausible property when deriving it in the
replicated system (with the replica number $n$ going to zero) \cite{Parisi04},
it can now be believed that the identities of replica equivalence are also
true for the standard site overlap in the finite dimensional Ising spin glass.
As both scenarios, RSB and droplet, fulfill replica equivalence, it can not,
however, be
used to distinguish them.\footnote{It has emerged as an exact statement
in the literature \cite{NeSt96,NeSt05}
that the distribution of the site overlaps for any
finite-dimensional EA model must be self-averaging --- $P(q)=P_J(q)$
--- in case of the infinite system. Putting together this thing with replica
equivalence, one can deduce a trivial overlap distribution, i.e.\ a droplet
scenario in any $d$. Marinari et al in Ref.\ \cite{Marinari_et_al}
argued that the $P_J(q)$ defined by Newman and Stein is different
from that appearing in the replica equivalence identities. 
Nevertheless, one might have
the feeling that this issue has not been completely settled by now.}

To review the huge amount of numerical work accumulated in the last
decades is impossible, the reader is referred
to Refs.\ \cite{MaPaRu98,Marinari_et_al} for extensive lists of papers.
Numerical simulations work on finite systems and try to extrapolate
to the infinite $N$ limit. Due to the extremely long relaxation times,
notwithstanding the huge evolution of computer capacities, only relatively
small systems have been simulated. That resulted in a fluctuating support
for the two scenarios. Ref.\ \cite{Marinari_et_al} considered almost
all topics of the above list, finding evidences for replica equivalence
{\em and\/} RSB. A very recent paper \cite{ultrametricity06}
for the three-dimensional $\pm J$ model in zero external field found
excellent agreement with ultrametricity --- the lattice size was
relatively large: $20^3$ ---, which is a unique characteristic of RSB.
The issue of the low-energy excitations (point 3) is rather controversial:
Refs.\ \cite{KrMa00,PalassiniYoung00} found system size $O(1)$ energy
excitations, but with $d_s<d$ suggesting that the link overlap distribution
is trivial, whereas the standard one is not (TNT picture). This
scenario, however, was ruled out by \cite{CoGiGiVe06}. Zero temperature
methods were also used to test the survival of the spin glass state
in a nonzero external field \cite{KrHoMaMaPa01}; the somewhat positive
answer obtained is in favour of RSB. On the other hand, the completely
different treatment in \cite{YoungKatzgraber04} concluded the lack of
a transition in field; though the magnetic field was not homogenous
in that investigation, but chosen from a random Gaussian distribution.
Finally point 5 is rather difficult to check numerically as the
exponents in (\ref{point 5}) are very close to each other, due to
the smallness of $\Theta/d$.

The replica method introduced by Edwards and Anderson \cite{EA}, is
a powerful, though admittedly unphysical, tool to handle the quenched
state by eliminating the inhomogenities caused by randomness. The effective
(replicated) Hamiltonian thus obtained can be
represented by a field theory using
common procedures known from the study of ordinary systems.The advantage of
field theory comes from its flexibility of application in any space dimension
$d$ and for a variable range of the interaction. Standard methods, like
systematic perturbation expansions and the renormalization group (RG), are
also available. In the first calculation by replica field theory of
Harris et al \cite{HaLuCh76} the fixed point value and the critical
exponents were obtained in leading order in $\epsilon=6-d$. In two subsequent
papers \cite {BrMo79,PytteRudnick79} the ordered phase below $T_c$, and
especially the stability of the RS state in finite $d$ was studied: the
mean field, AT-like, instability was found to survive perturbatively.
After Parisi's solution of the SK model appeared and proved to be (marginally)
stable \cite{DeKo83}, one-loop studies based on the ultrametric RSB ansatz
were published (see \cite{beyond}, and references therein), the severe
infrared problems were handled in the framework of a scaling theory
\cite{scaling_and_infrared}. In the meantime, a search for finding the
(presumably zero-temperature) fixed-point of a finite-dimensional AT line
took place \cite{BrRo,PRLcikk}.

In field theory, detection of a glassy phase consistent with the droplet
picture would mean finding a marginally stable RS state in zero external
field. The developments enumerated in the previous paragraph were against
droplet theory in finite dimensions. In Ref.\ \cite{BrMo79} a procedure
with temporarily breaking the replica symmetry by a special --- non-Parisi ---
type of RSB (the ``two-packet'' RSB), and eventually restoring RS led to a
massless RS phase. We will comment this result in Section \ref{Conclusion}.
The issue has been revived in two recent papers \cite{Moore,Cirano}.

Alike the solution of the fully connected SK model, field theory provides
results in the thermodynamic limit, i.e.\ for $N\to \infty$. A direct
comparison with the lattice EA model is, however, impossible due to the use
of the concept of the minimum length \cite{Ma}, i.e.\ averaging out
short-range fluctuations. As a result, the precise dependence of the
parameters of the field theoretical Lagrangian, the bare couplings, on the
control parameters (temperature, external magnetic field, etc.)
of the original model is lost. This fact may be extremely important when
trying to distinguish an AT line from a droplet-like phase \cite{droplet}:
both have the generic replica symmetry with a nonzero EA order parameter,
and even an infinite spin glass susceptibility (a zero replicon mass).
For that reason, we follow the ideas, introduced in Ref.\ \cite{droplet},
of using exact symmetry arguments to parametrize the bare couplings in
zero external field. Our arguments are based on the only assumption that
the Legendre transformed free energy is an invariant of the extra symmetry
of the high-temperature phase, see Eq.\ (\ref{invariance}) below.
An important result of the present paper is the
summation of the infinite terms which were argued in \cite{droplet}
to contribute to the replicon mass $2m_1$: there is a complete cancellation of
these terms leading to a negative mass, i.e.\ to an unstable RS phase for
$6<d<8$ and $d\lesssim 6$.

The outline of the paper is as follows: In Section \ref{symmetry} the generic
replica symmetric model is introduced, and the extra symmetry of the
replicated paramagnet of the high-temperature phase presented. Our basic
object, the replicon generator $g(v)$, is defined, and its properties are
derived by exact symmetry considerations. The full $g(v)$ 
is computed for the SK case in Section \ref{Mean field theory}.
The picture of independent droplets is used to derive the scaling form
of higher order susceptibilities in Section \ref{Droplet}. A somewhat
heuristic argument makes it possible to produce a scaling form of $g(v)$
for the droplet scenario when the replica number $n$ is nonzero.
Exact Ward identities, which are necessary to construct the free propagators,
are displayed in Section \ref{Ward},
whereas the one-loop calculation of $g(v)$ is presented
in Section \ref{one-loop}. The qualitatively different regimes --- i.e.
$d>8$, $6<d<8$ and $d\lesssim 6$ --- are discussed in separate subsections.
Section \ref{Conclusion} is for a summary of the results, and it also contains
our conclusions.
Notations, definitions and computational details
are left to the series of appendices,
amongst them the block diagonalization of a generic matrix with the
``two-packet'' RSB is in Appendix \ref{App5}.

\section{The symmetry of the spin glass field theory in zero
magnetic field}\label{symmetry}

The Ising spin glass on a hypercubic $d$-dimensional lattice
--- with the
quenched exchange interactions taken independently from a Gaussian distribution
with zero mean ---
can be represented
(see \cite{rscikk} and references therein)
by a replica field theory where the classical fields
$\phi^{\alpha\beta}$ have the {\em double\/} replica indices
$\alpha,\beta=1\dots n$, and $\phi^{\alpha\beta}=\phi^{\beta\alpha}$
with zero diagonal $\phi^{\alpha\alpha}$. In the spirit of the replica
trick \cite{EA}, the special limit $n\to 0$ of this $n(n-1)/2$ component
field theory provides the physical results. Unlike in the Parisi theory,
the spin glass phase is almost ergodic in the droplet scenario: the two
minimal free-energy phases are related by the global inversion of the spins
--- just like in a ferromagnet. In a replicated picture this gives
rise to an RS
theory with the generic Lagrangian (displayed here in momentum space):
\begin{equation}\label{L}
\mathcal{L}=\frac{1}{2}\sum_{{\mathbf p}}\bigg[
\Big(\frac{1}{2} p^2+\bar{m}_1\Big)\sum_{\alpha\beta}
\phi^{\alpha\beta}_{{\mathbf p}}\phi^{\alpha\beta}_{-{\mathbf p}}
+\bar{m}_2\sum_{\alpha\beta\gamma}
\phi^{\alpha\gamma}_{{\mathbf p}}\phi^{\beta\gamma}_{-{\mathbf p}}\\+
\bar{m}_3\sum_{\alpha\beta\gamma\delta}
\phi^{\alpha\beta}
_{{\mathbf p}}\phi^{\gamma\delta}_{-{\mathbf p}}
\bigg]
\,\,+\,\,\mathcal{L}^{\text{I}},
\end{equation}
with the interaction term containing all the higher order invariants
$I^{(k)}_j$ together with the single first order one, i.e.
\begin{equation}\label{LI}
\mathcal{L}^{\text{I}}=-N^{\frac{1}{2}}\,\bar{h}\,\sum_{\alpha\beta}
\phi^{\alpha\beta}_{{\mathbf p}=0}
-\sum_{k=3}
\,\frac{1}{k!\,N^{\frac{k}{2}-1}}\,
\sideset{}{'}\sum_{\mathbf {p_i}}\,\sum_j\,\bar{g}^{(k)}_jI^{(k)}_j.
\end{equation}
The number of spins $N$ goes to infinity in the thermodynamic limit,
its explicit indication
ensures that the bare couplings, $\bar{h}$ and the $\bar{g}^{(k)}_j$'s, are
independent of $N$. Momentum conservation is understood in the primed sum
over the momentums ${\mathbf p}_1\dots{\mathbf p}_k$. The $I^{(k)}_j$'s
are invariant under the global transformation ${\phi'}^{\alpha\beta}_{{\mathbf p}}=
\phi^{P_\alpha P_\beta}_{{\mathbf p}}$, where $P$ is any permutation of the $n$
replicas. The three invariants of the kinetic term ($k=2$)
and that of the linear one are explicitly
displayed in (\ref{L}) and (\ref{LI}), while the 8 cubic and 23 quartic ones are summarized
in Appendix \ref{App1}.\footnote{
\label{fn1}
The barred notation for the {\em bare\/} coupling
constans is used throughout the paper to distinguish them from their
{\em exact\/} counterparts. The choice of the set of coupling constants 
of the generic replica symmetric Lagrangian is obviously not unique:
couplings with the lowercase notation defined above are the coefficients
in front of the invariants. To make the presentation as clear as possible,
it will be necessary later to introduce another set with the capital letter
notation ($m_1$ vs.\ $M_1$, for instance); the linear relationships between
these two sets are displayed --- for the quadratic, cubic and quartic vertices
--- in Appendix
\ref{App2}.} Besides using $\bar{g}^{(k)}_j$ for a generic $k$, we
retain the commonly used notations for the low order couplings
$\bar{h}$, $-\bar{m}$'s, $\bar{w}$'s (for cubic) and $\bar{u}$'s (for quartic).
As it was pointed out in \cite{droplet}, an ambiguity in assigning
the bare coupling
constants to a given physical state occurs due to the freedom to
offset the zero-momentum fields,
$\phi^{\alpha\beta}_{{\mathbf p}=0}\longrightarrow
\phi^{\alpha\beta}_{{\mathbf p}=0}-\sqrt N\,\Phi$, leaving all the irreducible
vertices unaltered while the bare couplings changing. By this way, we can
always tune the one-point function to zero, hence avoiding 
the appearence of any ``tadpole'' insertions in a perturbative treatment.
We adopt this definition of the bare coupling constants in what follows.
(In this case, $\bar{h}$ is generally nonzero even if the ``physical''
magnetic field is switched off.)

The generic replica symmetric theory presented in the previous paragraph,
however, is not confined to the description of 
the droplet-like spin glass phase: the nonglassy phase in an external
magnetic field
(even above the transition temperature)
is also represented by the generic RS Lagrangian of
Eqs.\ (\ref{L},\ref{LI}). Still, as a spin glass is always an infinitely
correlated state, the criterium of a massless mode (the so called replicon
one) seemingly may distinguish it from the magnetized paramagnet.
At that point, however, we must realize that the boundary of a possible
RSB phase fits the same criterium: along the AT line we have replica
symmetric massless thermodynamic states.

The underlying problem we face when trying to distinguish an AT line
from the droplet-like spin glass phase is the indirect relationship
between the bare couplings of the field theory and the physical
control parameters (temperature and magnetic field, for instance).
To find the thermal route of the system, i.e.\ the dependence of the
bare couplings on the reduced temperature $\frac{T_c-T}{T_c}$ while
keeping the external magnetic field zero, we turn to exact symmetry
arguments. The ``true'' paramagnet, i.e.\ the high temperature phase in zero
magnetic field, has the extra symmetry that the Edwards-Anderson order
parameter $q$ is zero. (The magnetic field makes $q$ nonzero even in the
nonglassy phase.) As a result, those couplings in (\ref{L}) and (\ref{LI})
which belong to invariants with at least one replica occuring an odd
number of times (like $\bar{m_2}$, $\bar{m_3}$ and $\bar{h}$)
are exactly zero. In addition to the permutational invariance, this
results in the extra symmetry published in \cite{droplet},
and for the sake of self-containness, it is explained here
again: The $n$ replicas are divided arbitrarily into two groups with $p$
and $n-p$ replicas, $p$ being a free parameter between $0$ and $n$. The
Lagrangian is invariant under the global transformation
\begin{equation}\label{O} {\phi'}^{\alpha\beta}_{{\mathbf p}}=
\begin{cases}
\phi^{\alpha\beta}_{{\mathbf p}} & \text{for $\alpha$ and $\beta$ in the same group,}\\
-\phi^{\alpha\beta}_{{\mathbf p}} & \text{for $\alpha$ and $\beta$ in different groups.}
\end{cases}
\end{equation}
This transformation is obviously orthogonal in the $n(n-1)/2$-dimensional
space of the field components, with the diagonal transformation matrix
$O_{\alpha\beta,\gamma\delta}=(-1)^{\alpha\cap\beta+1}\,\,
\delta^{\text{Kr}}_{\alpha\beta,\gamma\delta}$. We introduced here the
overlap $\alpha\cap\beta$ defined as $\alpha\cap\beta=0$
($\alpha\cap\beta=1$) if $\alpha$ and $\beta$ are in different groups
(if $\alpha$ and $\beta$ are in the same group).

The symmetry of the Lagrangian is reflected in the invariance of the
Legendre-transformed free energy $F$:
\begin{equation}\label{invariance}
F(q'_{\alpha\beta})=F(q_{\alpha\beta}).
\end{equation}
The stationary condition $\frac{\partial F}{\partial q_{\alpha\beta}}
=0$ provides the symmetrical solution $q'_{\alpha\beta}=q_{\alpha\beta}
\equiv 0$ for the paramagnet, while in the droplet-like phase the extra
symmetry of (\ref{O}) is lost leading to $q_{\alpha\beta}\equiv q
\not= 0$ represented by the generic RS theory in (\ref{L}) and (\ref{LI}).
In this way, we can understand the paramagnet to spin glass transition as a
symmetry breaking one.

In \cite{droplet} the invariance property of (\ref{invariance}) was combined
with the freedom to choose the {\em continuous\/} parameter $p$ of the
transformation to derive a set of Ward-like identities between exact
irreducible vertex functions (derivatives of $F$ evaluated at stationarity).
Here we focus on the stability of the low-temperature RS phase in zero
magnetic field, and for that reason we define the generator function $g(v)$
of replicon-like vertices as follows:
\begin{equation}\label{g}
g(v)=\sum_{k=2}^{\infty}\,\frac{1}{k!}\,G^{(k+1)}v^k,
\end{equation}
where
\footnote{\label{fn2}Throughout the paper, we use different notations for
summations over distinct pairs
of replicas, $\sum\limits_{(\alpha\beta)}=\sum\limits_{\alpha<\beta}$,
and unrestricted sums, $\sum\limits_{\alpha\beta}=
\sum\limits_{\alpha}\sum\limits_{\beta}$, as in the invariants in
Eq.\ (\ref{L}).}
\begin{equation}\label{G^(k+1)}
G^{(k+1)}=
\sideset{}{'}\sum_{(\alpha_1\beta_1),\dots ,(\alpha_k\beta_k)}
G^{(k+1)}_{\alpha\beta,\alpha_1\beta_1,\dots ,\alpha_k\beta_k},\quad
k=1,2,\dots;\quad\text{and by definition}\quad G^{(1)}\equiv
G^{(1)}_{\alpha\beta}.
\end{equation}
The prime above is to indicate the restriction $\alpha_i\cap\beta_i=0$
for all $i=1,\dots,k$ in the summations, and we choose $\alpha\cap\beta=0$
too, hence $G^{(k+1)}$ no more depends on $\alpha$ and $\beta$ in an RS state
(it does, however, on $n$ and $p$).
The generic definition of the {\em exact\/} vertices is
\begin{equation}\label{vertices}
-G^{(k)}_{\alpha_1\beta_1,\dots ,\alpha_k\beta_k}=
\frac{\partial^k (F/N)}{\partial q_{\alpha_1\beta_1}\dots\partial q_{\alpha_k\beta_k}}
\end{equation}
evaluated at an RS stationary point. For $k=2,3$ and $4$ we insist
--- for traditional reasons --- to use the
notations $-M$, $W$ and $U$, respectively,
similarly to their lowercase counterparts, see the note below
Eqs.\ (\ref{L},\ref{LI}).
The  vertices defined in (\ref{vertices}) take only a limited
number of different values, expressing in this way the RS symmetry;
in fact they may be numbered in accord with the numbering of the
set with the lowercase notation.
All of the relevant definitions are left to Appendix \ref{App1}, while
the linear relationships between the two sets are displayed in Appendix
\ref{App2}
for the quadratic ($k=2$), cubic ($k=3$) and quartic ($k=4$) cases.
(See also the footnote \ref{fn1}.)

Differentiating both sides of Eq.\ (\ref{invariance}), and exploiting the
orthogonality and diagonality of the transformation in (\ref{O}),
the following rule is easily derived for the transformation of the vertices:
\begin{equation}\label{vertex trafo}
G^{(k)\,'}_{\alpha_1\beta_1,\dots ,\alpha_k\beta_k}=
O_{\alpha_1\beta_1,\alpha_1\beta_1}\dots O_{\alpha_k\beta_k,\alpha_k\beta_k}
\,G^{(k)}_{\alpha_1\beta_1,\dots ,\alpha_k\beta_k},
\end{equation}
inferring immediately
\begin{equation}\label{G^(k+1)trafo}
G^{(k+1)\,'}=(-1)^{k+1}\,G^{(k+1)}.
\end{equation}
As it was explained in \cite{droplet}, the transformation $q_{\alpha\beta}
\rightarrow q'_{\alpha\beta}$ becomes infinitesimal for $q$,
$n$, $p\ll 1$, hence
the left-hand side of (\ref{G^(k+1)trafo}) can be Taylor-expanded around
the RS state $q_{\alpha\beta}\equiv q$. From (\ref{G^(k+1)}) and
(\ref{vertices}), and using $q'_{\alpha\beta}-q_{\alpha\beta}=0$ unless
$\alpha\cap\beta=0$, when it is $-2q$, we can easily derive:
\begin{equation}\label{Taylor}
G^{(k+1)\,'}=\sum_{l=0}^\infty \,\frac{1}{l!}\,\,G^{(k+1+l)}\,(-2q)^l.
\end{equation}
Comparing (\ref{G^(k+1)trafo}) and (\ref{Taylor}),
\[
\left[(-1)^{k+1}-1\right]\,G^{(k+1)}=
\sum_{l=1}^\infty \,\frac{1}{l!}\,\,G^{(k+1+l)}\,(-2q)^l
\]
is concluded, valid for $k=0,1,\dots$. It is useful to analyze the following
cases separately:
\begin{itemize}
\item
For $k=0$, using (\ref{g}), $-2\,G^{(1)}= G^{(2)}\times(-2q)+g(-2q)$
is obtained. Stationarity renders the left-hand side zero. Here
we give some details of the derivation of $G^{(2)}$ in terms of the masses,
intending in this way to facilitate the reader to check more complicated
formulae for $G^{(3)}$ and $G^{(4)}$ displayed in Appendix \ref{App3}:
\begin{equation}\label{G^(2)}
-G^{(2)}=
\sideset{}{'}\sum_{(\gamma\delta)}\,M_{\alpha\beta,\gamma\delta}=
M_1+\left[(p-1)+(n-p-1)\right]M_2+(p-1)(n-p-1)M_3;
\end{equation}
for the notations used here see Appendix
\ref{App1}. This formula and the similar
ones displayed in Appendix \ref{App3}
are much simpler when expressed in terms of the lower
case vertices, though the derivation uses evidently the upper case set.
By Eq.\ (\ref{m and w}) of Appendix
\ref{App2}, we arrive to the important result
\begin{equation}\label{k=0}
2m_1+nm_2+4p(n-p)m_3=\frac{g(v)}{v}\quad\text{evaluated at}\quad
v=-2q.
\end{equation}
\item
For $k=1$ the right-hand side is just $g'(-2q)$, i.e.\ 
\begin{equation}\label{k=1}
g'(v)=0\quad\text{at}\quad
v=-2q.
\end{equation}
\item
The generic case, $k\ge 2$, is easily understood by observing that the
right-hand side is $g^{(k)}(-2q)-G^{(k+1)}$, while $G^{(k+1)}=g^{(k)}(0)$
providing
\begin{equation}\label{generic k}
(-1)^{k+1}\,g^{(k)}(v=0)=g^{(k)}(v=-2q).
\end{equation}
\end{itemize}

The above conditions are serious restrictions for the RS field theory
representing a spin glass phase in zero magnetic field. For a given $q$
(which is used here --- instead of the reduced temperature ---
to parametrize the spin glass phase below $T_c$), the generator $g(v)$
must have an extremum at $v=-2q$, Eq.\ (\ref{k=1}), while the higher
derivatives satisfy (\ref{generic k}). As $2m_1$ is the dangerous
(so called ``replicon'') mass responsible for the instability of the RS
state \cite{AT}, stability requires $g(v=-2q)\le 0$ in the limit $n\to 0$.
(Equality means marginal stability, i.e.\ a massless spin glass phase,
just what droplet theory suggests \cite{BrMo86,FiHu88}.) These results were
derived by generic symmetry arguments, hence they must be equally valid
for systems described by mean field theory
and for short-range
spin glasses in physical dimensions. All these cases are reviewed
in subsequent sections.

\section{Calculation of the replicon generator in mean field theory}
\label{Mean field theory}

An explicit expression for $F(q_{\alpha\beta})$ is available in mean
field theory, working it out either on the fully connected lattice
(Sherrington-Kirkpatrick model \cite{SK}) or as the long-range limit on the
$d$-dimensional hypercubic lattice \cite{rscikk}:
\begin{equation}\label{SK F}
\frac{1}{N}F=\frac{1}{2\Theta}\sum_{(\alpha\beta)}q_{\alpha\beta}^2
-\ln \underset{\{S^{\alpha}\}}{\mathrm {Tr}}\exp \bigg(\sum_{(\alpha\beta)}
q_{\alpha\beta}S^{\alpha}S^{\beta}\bigg).
\end{equation}
As we are in zero magnetic field, the only control parameter is
$\Theta\sim (kT)^{-2}$. It is now useful to introduce the modified
Legendre-transformed free energy $\tilde F$ by
\begin{equation}\label{modF}
\tilde F(v_{\alpha\beta})=F(q+v_{\alpha\beta}),
\end{equation}
$q$ being the RS stationary point of $F$. Obviously $\tilde F$ is stationary
for $v_{\alpha\beta}\equiv 0$, and all the vertices in Eq.\ (\ref{vertices})
can be computed equally well substituting $F$ by $\tilde F$ and the $q$'s
by $v$'s. (The invariance property of (\ref{invariance}) is, however, lost
for $\tilde F$.) After Taylor-expanding $\tilde F$ and making the restriction
$v_{\alpha\beta}=v$ for $\alpha\cap\beta=0$, otherwise it is zero, we can
express $g(v)$ by the now one-variable function $\tilde F(v)$:
\[
g(v)=-vG^{(2)}-\frac{1}{p(n-p)}\,\frac{d}{dv}
(\tilde F/N).
\]
The second term above is the harder to compute. As it is common in mean field
replica calculations, the Gaussian integral representation method can
eliminate the spin traces in (\ref{SK F}) providing for $p(n-p)\to 0$:
\begin{multline}\label{mean field g}
g(v)=\\[8pt]
\frac{\int \mathcal{D}_y \int \mathcal{D}_x \int \mathcal{D}_{x'}\,
\tanh(\sqrt{q+v}\,y+\sqrt{-v}\,x)\tanh(\sqrt{q+v}\,y+\sqrt{-v}\,x')\cosh^n
(\sqrt{q+v}\,y+\sqrt{-v}\,x)}
{\int \mathcal{D}_y \int \mathcal{D}_x \int \mathcal{D}_{x'}\,
\cosh^n(\sqrt{q+v}\,y+\sqrt{-v}\,x)}\\[8pt]
-\overline{\tanh^2(\sqrt{q}\,y)}-\left[1+(n-2)\,\,\overline{\tanh^2(\sqrt{q}\,y)}
-(n-1)\,\,\overline{\tanh^4(\sqrt{q}\,y)}\right]v.
\end{multline}
The shorthand notation 
\[
\overline{\tanh^k(\sqrt{q}\,y)}\equiv
\frac{\int \mathcal{D}_y \,\tanh^k(\sqrt{q}\,y)\cosh^n(\sqrt{q}\,y)}
{\int \mathcal{D}_y \cosh^n(\sqrt{q}\,y)}\quad\text{with}\quad
\int \mathcal{D}_y\equiv\int\frac{dy}{\sqrt{2\pi}}\,e^{-\frac{1}{2}y^2}
\]
was used here. The explicit temperature dependence has been
eliminated by the stationary condition $\Theta^{-1}q=
\overline{\tanh^2(\sqrt{q}\,y)}$, and $g(v)$ is thus 
parametrized entirely by the
order parameter $q$.

We are now in the position to check the generic properties of $g(v)$
deduced in the preceeding section. It is relatively easy to see that
$g(v)$ is quadratic for small $v$. Although it is somewhat harder,
a straightforward application of the following two lemmas%
\footnote{The first one can be easily proven by a sequence of partial
integration, while a change of variables to $z$, $z'$ and
$y'=y/\sqrt{-v}$ in the second lemma
and setting $v=-2q$ renders the integral over $y'$
proportional to $\delta(z+z')$, leading us immediately to the end of
the proof.} can prove Eqs.\ (\ref{k=1}) and (\ref{generic k}):
\begin{enumerate}
\item
\[
\frac{d}{dv}\,
\int \mathcal{D}_y \int \mathcal{D}_x \int \mathcal{D}_{x'}\,f_1(z)f_2(z')=
\int \mathcal{D}_y \int \mathcal{D}_x \int \mathcal{D}_{x'}\,
f'_1(z)f'_2(z')
\]
and
\item
\[
\int \mathcal{D}_y \int \mathcal{D}_x \int \mathcal{D}_{x'}\,f_1(z)f_2(z')=
\int \mathcal{D}_z\, f_1(\sqrt{q}\,z)f_2(-\sqrt{q}\,z)
\quad \text{for}\quad v=-2q.
\]
\end{enumerate}
$f_1$ and $f_2$ above are two arbitrary functions,
while $z=\sqrt{q+v}\,y+\sqrt{-v}\,x$ and $z'=\sqrt{q+v}\,y+\sqrt{-v}\,x'$.

Near $T_c$, $g(v)$ gets into the simple scaling form $g(v)=q^3\,\hat g
(v/q)$; the scaling function results from expanding the integrands
of (\ref{mean field g}) for $q\ll 1$ and $v/q=O(1)$:
\begin{equation}\label{ghat}
\hat g(x)=-\frac{3n-2}{3}x^2(3+x),\qquad p(n-p)\to 0.
\end{equation}
The following remarks are appropriate here:
\begin{itemize}
\item $\hat g(-2)>0$, showing by (\ref{k=0}) the instability of the RS
mean field spin glass phase just below $T_c$.
\item The leading, or scaling, term of $g(v)$
--- and only that --- satisfies the following
condition:
\begin{equation}\label{diff g}
g(v=-2q)=\frac{2}{3}\,q^2\,g''(v=0).
\end{equation}
We shall see later that this property of the replicon generator is very
robust, as it persists for the short-range system below the upper
critical dimension 6.%
\footnote{To be more precise, the {\em one-loop\/} calculations in
Section \ref{one-loop} will show that Eq.\ (\ref{diff g}) is true
for {\em any\/} $n$ when $d>8$, whereas for $6<d<8$ and $d\lesssim 6$ it is
restricted to the case $n=0$.}
On the other hand, a simple one term Ward-identity
follows from (\ref{diff g}) for the replicon mass,
as one can see from (\ref{k=0}) and (\ref{diff g}):
\begin{equation}\label{m1 vs w2}
2m_1=\frac{g(v=-2q)}{(-2q)}=-\frac{1}{3}q\,G^{(3)}=-\frac{2}{3}q\,w_2,
\qquad p,n\to 0.
\end{equation}
[The expression for $G^{(3)}$ in terms of the $w$'s is displayed in 
Appendix \ref{App3}, see Eq.\ (\ref{W0}).] This formula is valid whenever the leading
contribution to $g(v)$ for $n\to 0$ and $q\to 0$ satisfies (\ref{diff g}).
As it connects the replicon mass ($2m_1$) and the replicon-like cubic
vertex ($w_2$), it is obviously important in testing stability. We shall
therefore return to it later.
\item The leading term of $g(v)$ in Eq.\ (\ref{ghat}) can be deduced directly
from the truncation of the infinite series (\ref{g}) as
\begin{equation}\label{truncated g}
g(v)=\frac{1}{2}G^{(3)}v^2+\frac{1}{6}G^{(4)}v^3.
\end{equation}
This can be seen by a direct evaluation of $G^{(3)}$ and $G^{(4)}$ from
the definitions (\ref{G^(k+1)}), (\ref{vertices}) and using the mean field
free energy (\ref{SK F}). This truncated form of the scaling part of the
replicon generator is, however, more general, as it remains valid for the
finite-dimensional system when $d>8$ (see later). Starting from
(\ref{truncated g}), Eq.\ (\ref{diff g}) trivially follows, together
with all of its consequences.
\end{itemize}

From the generic properties of $g$ derived in the preceding section and
from the mean field analysis above, one may get the feeling that the
replicon mass [which is proportional to $g(v=-2q)$] is either negative
or positive, but it is difficult to imagine how marginal stability may
emerge. This, however, belongs to the essence of the droplet picture,
and in the next subsection we try to construct $g(v)$
--- or rather its scaling term ---using the ideas
of the theory of free droplets.

\section{The replicon generator in the droplet picture}

\label{Droplet}
The droplet picture is a low-temperature scaling theory developed for
the original $d$-dimensional lattice system
\cite{McMillan,BrMo86,FiHu86,FiHu88}. As a basic
inference of this theory, the spin glass state is now restricted to zero
external magnetic field, and it is in fact a critical surface attracted by
a zero-temperature fixed point, immediately suggesting that temperature
is now a (dangerously) irrelevant parameter. The whole phase below $T_c$
is massless, as at least one correlation length, namely that of the
replicon mode, is infinite. Somewhat surprisingly, there is a single
independent exponent belonging to the zero-temperature fixed point
--- the stiffness exponent $\theta$ ---, and scaling of relevant and
irrelevant variables, and also of several observables (correlation
functions, for instance), can be described by means of it.

Translating the results of the droplet theory to the language of replicated
field theory is straightforward by using the Gaussian integral representation,
called Hubbard-Stratonovich transformation, and then eliminating short ranged
fluctuations by the introduction of effective coupling constants and truncation
of the momentum-dependent mass term (see Ref.\ \cite{rscikk}); a Lagrangian
like that in Eqs.\ (\ref{L}) and (\ref{LI}) is obtained. As a short cut to
translate different quantities, we propose that whenever a pair of replicated
Ising spins $S_i^{\alpha}S_i^{\beta}$ occurs ($i$ being a lattice site),
it should be replaced by the field ${\phi}_i^{\alpha\beta}$; in this
way, Ising averages are turned to averages with the measure
$e^{-\mathcal L}$. 
As a pedagogical example,
let us consider the momentum-dependent Edwards-Anderson susceptibility
defined by
\[
\chi_{\text{EA}}(p)=\frac{1}{N}\sum_{ij}e^{i{\mathbf p}
({\mathbf r}_j-{\mathbf r}_i)}\,\, \overline{(\langle S_iS_j\rangle
-\langle S_i\rangle\langle S_j\rangle)^2},
\]
where $\langle \dots \rangle$ and $\overline{\rule{0pt}{6pt}\dots}$ mean thermal
average with the Ising measure and quenched average over the disorder,
respectively. Applying the replica trick and introducing the fields, we get
\begin{multline*}
\chi_{\text{EA}}(p)=\frac{1}{N}\sum_{ij}e^{i{\mathbf p}
({\mathbf r}_j-{\mathbf r}_i)}
\big(\langle S_i^{\alpha}S_i^{\beta} S_j^{\alpha}S_j^{\beta}
-2S_i^{\alpha} S_i^{\gamma}S_j^{\beta} S_j^{\gamma}+
 S_i^{\alpha}S_i^{\beta}S_j^{\gamma}S_j^{\delta}\rangle\big)\\
 \longrightarrow\qquad
 (kT)^b\,\big(\langle \phi_{\mathbf p}^{\alpha\beta}
 \phi_{-\mathbf p}^{\alpha\beta}\rangle-2\langle \phi_{\mathbf p}^{\alpha\gamma}
 \phi_{-\mathbf p}^{\beta\gamma}\rangle +\langle \phi_{\mathbf p}^{\alpha\beta}
 \phi_{-\mathbf p}^{\gamma\delta}\rangle \big)\equiv (kT)^b\,C^{(2)}(p);
\end{multline*}
$C^{(2)}(p)$ being the replicon component of the 2-point connected correlation
function in the replicated field theory. (To avoid complicated notations,
we stick to $\langle \dots \rangle$ for thermal averages with different
measures: in the first line it is taken with the effective replicated Ising
Hamiltonian resulting from the replica trick, whereas a field-theoretic
average with the weight $e^{-\mathcal{L}}$ is understood in the second one.)
The emergence of the factor $(kT)^b$ above results from the transformation.
Similar temperature-dependent factors with some simple exponents always
appear whenever spin averages are represented by field averages. Near $T_c$
they are absolutely irrelevant, we must, however, keep them at low
temperature to ensure the correct temperature scaling. (Nevertheless,
we shall see that the specific values of 
$b$ and also $a$, which is introduced later,
will not influence our basic conclusion.) 
We know from Refs.\ \cite{FiHu86,BrMo86,FiHu88} that $\chi_{\text{EA}}(p)\sim
(kT)\,p^{-d+\theta}$ yielding
\begin{equation}\label{C^(2)}
C^{(2)}(p)\sim (kT)^{1-b}\,p^{-d+\theta}.
\end{equation}

We can apply the scaling formula
\[
C^{(2)}(p)\cong {\mu_1}^{\frac{-2+\eta_{\text{R}}}{\lambda_1}}
\,\,{\tilde C}^{(2)}
\bigg(\frac{\mu_2}{{\mu_1}^{\lambda_2/\lambda_1}},\frac{p}{{\mu_1}^{1/\lambda_1}}
\bigg)
\]
around the zero-temperature fixed point. $\mu_1$ is a relevant scaling field
vanishing in the spin glass phase
(e.g.\ the external magnetic field) with $\lambda_1>0$ exponent, while
$\mu_2\sim (kT)^a$ is the dangerously irrelevant one with $\lambda_2<0$.
The scaling index of
the replicon correlation function is $-2+\eta_{\text{R}}$, and the reader is
reminded that, in principle, there are two different $\eta$'s ($\eta_{\text{R}}$
and $\eta_{\text{L}}$) at a generic RS fixed point when $n=0$. On the one hand,
we can now exploit the fact that $C^{(2)}$ must be independent of $\mu_1$ when
it goes to zero, and the correct momentum-dependence must be ensured. On the
other hand, droplet theory suggests $\lambda_2=-a\,\theta$. From these, we can
conclude:
\begin{equation}\label{C^(2) scaling}
C^{(2)}(p)\cong (kT)^{\frac{2-\eta_{\text{R}}+\theta-d}{\theta}}\,
p^{-d+\theta}.
\end{equation}
Comparing (\ref{C^(2)}) and (\ref{C^(2) scaling}), we get
\begin{equation}\label{etaR}
\eta_{\text{R}}=2-d+b\,\theta.
\end{equation}
The replicon mass [$2m_1$; see the table for the masses in Appendix \ref{App1}
and (\ref{m and w})] is --- by Dyson's equation \cite{Amit} ---
just the zero momentum limit of
$C^{(2)}(p)^{-1}$, and therefore obviously zero for any $T$ in the spin glass
phase.

The momentum-dependence of the higher order correlation functions
$C^{(k)}(p)$, $k>2$, can be deduced by applying the single droplet exitation
picture of Fisher and Huse \cite{FiHu86,FiHu88} for generalized Edwards-Anderson
susceptibilities. As an example, for $k=3$ we have:
\begin{multline*}
\frac{1}{N}\sum_{ijk}e^{i
({\mathbf p}_1{\mathbf r}_i+{\mathbf p}_2{\mathbf r}_j
+{\mathbf p}_3{\mathbf r}_k)}\,\, \overline{(\langle S_iS_jS_k\rangle
-\langle S_i\rangle\langle S_jS_k\rangle
-\langle S_j\rangle\langle S_iS_k\rangle
-\langle S_k\rangle\langle S_iS_j\rangle
+2\langle S_i\rangle\langle S_j\rangle\langle S_k\rangle)^2}\\
\longrightarrow\qquad \delta_{{\mathbf p}_1+{\mathbf p}_2+{\mathbf p}_3}\,
(kT)^{b'}\,C^{(3)}(p),\quad b'=\frac{3}{2}b,
\end{multline*}
where momentum conservation is explicitly shown, and only a single $p$
scale is considered. $C^{(3)}(p)$ has the structure of $2w_2$, see
(\ref{m and w}) and the table for the cubic vertices in Appendix \ref{App1},
and it is fully replicon-like \cite{rscikk}. The lengthy formula for
$C^{(3)}(p)$ is omitted here; the reader can reproduce it by replacing a
cubic vertex $W_{\alpha\beta,\gamma\delta,\mu\nu}$ in $2w_2$ by the average
$\sqrt{N}\,\langle \phi_{{\mathbf p}_1}^{\alpha\beta}\phi_{{\mathbf p}_2}^{\gamma\delta}
\phi_{{\mathbf p}_3}^{\mu\nu}\rangle$,%
\footnote{\label{N}The $\sqrt{N}$ factor here is to make $C^{(3)}(p)$ become
independent of $N$, i.e.\ of order unity.
For the same reason, we define $C^{(k)}(p)$ with the prefactor
$N^{\frac{k}{2}-1}$.}
and also symmetrizing
with respect to the momentums. In the spirit of the droplet theory, we can
conclude the momentum-dependence of the cubic and even higher order
replicon correlators as
\[
C^{(k)}(p)\sim p^{\theta-(k-1)d},\quad k=2,3,\dots\quad.
\]

It follows from simple scaling arguments that the replicon correlators,
which are by definition independent of $N$ (see footnote \ref{N}),
have the scaling index $d+\frac{k}{2}(\eta_{\text{R}}-2-d)$, and for
$\mu_2\ll{\mu_1}^{\lambda_2/\lambda_1}$ we have
\begin{equation}\label{C^(k) scaling}
C^{(k)}(p)\cong {\mu_1}^{[2d+k(\eta_{\text{R}}-2-d)]/2\lambda_1}
\,\,\big(\mu_2/{\mu_1}^{\lambda_2/\lambda_1}\big)^{\kappa_k}\,\,
{\tilde C}^{(k)}\big(p/{\mu_1}^{1/\lambda_1}\big),
\end{equation}
where the one-variable scaling function ${\tilde C}^{(k)}(x)\sim
x^{\theta-(k-1)d}$ for $x\to \infty$, whereas --- as the replicon correlation
length is finite for $\mu_1\not =0$ --- it goes to a constant in the opposite
limit. $\kappa_k$ is nonzero, expressing the dangerously invariant nature
of $\mu_2$, and it can be computed by taking the limit $\mu_1\to 0$,
while $\mu_2$ and $p$ finite, of $C^{(k)}(p)$ in (\ref{C^(k) scaling}):
\begin{equation}\label{kappa}
\kappa_k\,\lambda_2=k\,\bigg(\frac{\eta_{\text{R}}}{2}-1+\frac{d}{2}\bigg)
-\theta=
\bigg(\frac{k}{2}b-1\bigg)\theta,
\end{equation}
where the last equality follows from (\ref{etaR}). We shall need the scaling
of the zero-momentum correlation functions when approaching the spin glass
state for a given temperature. It can be easily deduced from
(\ref{C^(k) scaling}) and (\ref{kappa}):
\begin{equation}\label{relevant scaling}
C^{(k)}(p=0)\sim \mu_1^{[\theta-(k-1)d]/\lambda_1},\quad kT>0\quad \text{fixed}.
\end{equation}

It is pointed out in Appendix \ref{App3} that $G^{(k)}$, Eqs.\ (\ref{G^(k+1)})
and (\ref{vertices}), is a fully replicon vertex in the spin glass limit,
and therefore --- applying the usual connection between vertices
and connected correlation
functions \cite{Amit} --- it scales like $C^{(k)}(p=0)\times C^{(2)}(p=0)^{-k}$.
Using (\ref{relevant scaling}), this gives
\begin{equation}\label{R vertex scaling}
G^{(k)}\sim \mu_1^{[d-(k-1)\theta]/\lambda_1}
,\quad kT>0\quad \text{fixed}.
\end{equation}
This equation, which is valid for $n=0$, shows that $g(v)$ of the
spin glass in the droplet theory is highly singular
for $\mu_1\to 0$, i.e.\ for zero magnetic field. This is in full contrast
with mean field theory, see Eq.\ (\ref{ghat}). Nevertheless,
(\ref{R vertex scaling}) can be used to guess the $n$-dependence of $g(v)$
in the droplet picture in the following way:
Although Eq.\ (\ref{R vertex scaling}) was deduced by
renormalization group arguments, it merely expresses the assumption that an
infinitezimal magnetic field drastically destroys the droplet picture
by introducing a finite correlation length. Regarding the common observations
(see \cite{rscikk} for instance) that the role of the replica number is
alike, i.e.\ it makes the replicon mode more massive, we can assume tentatively
the following form for the $n$-dependence of $g(v)$ in zero magnetic field:
\begin{equation}\label{n scaling}
g(v)=n^{d\rho/\theta}\,\,\tilde{g}(v/n^{\rho}).
\end{equation}
This equation is obtained by the formal replacement of $\mu_1$ by $n$ in
(\ref{R vertex scaling}), and then substituting $G^{(k)}$ into the
definition of $g(v)$ in Eq.\ (\ref{g}). A new independent exponent $\rho$
was introduced here, and
--- although not explicitly indicated ---
the function $\tilde{g}$ depends on $T$ too. We shall return to this formula
later, extending it to the critical scaling region around $T_c$. 

\section{Ward identities and the free propagator for $T\lesssim T_c$}
\label{Ward}

Some of the identities between exact vertices presented in
Ref.\ \cite{droplet} follow directly from the properties of $g(v)$.
Eqs.\ (\ref{g}) and (\ref{k=0})
provide:
\begin{equation}\label{Ward1}
2m_1+nm_2+4p(n-p)m_3=-\Big(G^{(3)}-\frac{2}{3}G^{(4)}\,q+\dots\Big)\,q,
\end{equation}
whereas it follows from both (\ref{k=1}) and (\ref{generic k}):
\begin{equation}\label{Ward2}
G^{(3)}=G^{(4)}\,q+\dots\quad.
\end{equation}

One more expression arises from the transformation property of a mixed
mass component
\[
M'_{\alpha\beta,\gamma\delta}=-M_{\alpha\beta,\gamma\delta},\qquad
\alpha\cap\beta=1\quad\text{and}\quad\gamma\cap\delta=0
\]
[see (\ref{vertex trafo})], summing over $\gamma\cap\delta=0$ and expanding
the left-hand side. We get for $p\to 0$
\begin{equation}\label{Ward3}
m_2=-\big[(w_1+\frac{1}{3}w_3)+n\,\big(\frac{1}{3}w_5+w_6\big)\big]\,q
+\dots\quad.
\end{equation}
This equation has also been displayed in Ref.\ \cite{droplet}. Everywhere
in the above formulae, the dots have the obvious meaning of higher order
vertices multiplied by the appropriate power of $q$.

Eqs.\ (\ref{Ward1}), (\ref{Ward2}) and (\ref{Ward3}) are now used to
construct the bare propagators for a one-loop calculation of the leading
scaling term of $g(v)$. The following two regimes are separately treated:
\begin{itemize}
\item In the perturbative regime ($d>6$) the system is defined by the set
of bare couplings compatible with the symmetry of the high-temperature
phase: $\bar{w}_1$; $\bar{u}_1$, $\bar{u}_2$, $\bar{u}_3$, $\bar{u}_4$;
\ldots\ . The zero-loop limit of (\ref{Ward1}) and (\ref{Ward2}),
using also (\ref{U0}) of Appendix \ref{App3}, gives
\[
2\bar{m}_1+n\,\bar{m}_2=-\frac{2}{3}\bar{u}_2\,q^2,\qquad\qquad
2\bar{m}_3=-\frac{1}{3}(\bar{u}_1+2\bar{u}_4)\,q^2;
\]
whereas (\ref{Ward3}) provides 
\[
\bar{m}_2=-\bar{w}_1\,q.
\]
Neglecting the $q^2$ term in the scaling limit, the free propagators assume the
diagonalized form \cite{rscikk}
\begin{equation}\label{free propagators}
\frac{1}{p^2+r_s},\quad \text{with}\quad r_{\text{R}}=n(\bar{w}_1q),
\quad \quad r_{\text{A}}=2(\bar{w}_1q)\quad \text{and}
\quad r_{\text{L}}=(2-n)(\bar{w}_1q).
\end{equation}
[The multiplicities are: $n(n-3)/2$ for R (replicon), $n-1$ for A
(anomalous) and 1 for L (longitudinal); for the details, see \cite{rscikk}.]
\item For $d<6$ we enter the nonperturbative regime. Nevertheless, for
$6-d=\epsilon\ll 1$ the fixed point $\bar{w}^*_1$ is small, and the
perturbative renormalization group results of Ref.\ \cite{Iveta} are
usable. By the scaling analysis in \cite{droplet} we have shown that
$G^{(k)}\,q^{k-2}\sim \bar{w}^{*2}_1\,(\bar{w}^*_1q)^{\frac{\gamma}{\beta}}$,
with $\bar{w}^{*2}_1=O(\epsilon)$ and ${\gamma}/{\beta}=1+O(\epsilon)$.
The right-hand side of (\ref{Ward1}) and (\ref{Ward3}) are of order
$\epsilon$, with the only exception of $-w_1\,q$ in Eq.\ (\ref{Ward3}).
This results in the bare masses
\[
2\bar{m}_1+n\,\bar{m}_2=O[\epsilon\,(\bar{w}^*_1q)],
\qquad \bar{m}_3=O[\epsilon\,(\bar{w}^*_1q)]
\qquad \text{and}\qquad \bar{m}_2=-(\bar{w}^*_1q)\,[1+O(\epsilon)].
\]
After replacing $\bar{w}_1$ by $\bar{w}^*_1$, the free propagators remain the
same as in (\ref{free propagators}). 
\end{itemize}
In conclusion, in the one-loop calculation of the next
section we can use as bare masses:
\begin{equation}\label{bare masses}
2\bar{m}_1+n\,\bar{m}_2=0,\qquad \bar{m}_3=0
\qquad \text{and}\qquad \bar{m}_2=
\begin{cases}
-\bar{w}_1q & \text{for $d>6$} \\
-\bar{w}^*_1q & \text{for $d<6$.}
\end{cases}
\end{equation}

\section{The one-loop calculation of $g(v)$}
\label{one-loop}

Our basic formula for $g(v)$
can be derived by a Legendre-transform method; leaving the technical details
to Appendix \ref{App4}, we only quote the result here:
\begin{equation}\label{basic}
\begin{aligned}
g(v)&=\bar g(v)+\bar{H}+v \sideset{}{'}\sum_{(\gamma\delta)}
\big({M}_{\alpha\beta,\gamma\delta}-
{\bar{M}}_{\alpha\beta,\gamma\delta}\big)
+
\frac{1}{p(n-p)}\,\,\frac{d}{dv}\bigg[ \frac{1}{N}\ln \tilde Z\bigg],
\quad \text{with}\\[8pt]
\bar g(v)&=\frac{1}{2}\,v^2 \sideset{}{'}\sum_{(\gamma\delta),(\mu\nu)}
\bar{W}_{\alpha\beta,\gamma\delta,\mu\nu}+\frac{1}{6}\,v^3
\sideset{}{'}\sum_{(\gamma\delta),(\mu\nu),(\rho\omega)}
\bar{U}_{\alpha\beta,\gamma\delta,\mu\nu,\rho\omega}+\dots,
\quad \text{and} \quad \alpha\cap\beta=0.
\end{aligned}
\end{equation}
[Alike in (\ref{G^(k+1)}), the primed summation means the constraint that
the replicas in a summing pair belong to different groups.]
The second and third terms ensure that $g(0)=g'(0)=0$, as it must be by the
definition in (\ref{g}), whereas $\bar g(v)$ is the
replicon generator built up from the bare couplings. Due to the choice
$v_{\alpha\beta}=v$ for $\alpha\cap\beta=0$, and otherwise zero,
the partition function $\tilde{Z}$ of the system with the couplings
in (\ref{tilded couplings}) depends now on the single variable $v$.
Obviously $\ln\tilde{Z}$ has the ``two-packet'' RSB structure, nevertheless
it serves as the generator of the replicon-like vertices of the RS system.
Although
Eq.\ (\ref{basic}) will be used in a first order perturbative calculation
in this section, it is quite generic, and may be useful even in a
nonperturbative treatment.

The one-loop contribution to the last term of (\ref{basic}) is
\begin{equation}\label{g1}
g_1(v)=-\frac{1}{2p(n-p)}\,\frac{1}{N}\sum_{\mathbf p}\sum_j
\frac{1}{p^2+\lambda_j}\,\frac{\partial \lambda_j}{\partial v}\,\,,
\end{equation}
where $\lambda_j$ is an eigenvalue of $\tilde M$,
the bare mass of $\tilde{\mathcal L}$ displayed in (\ref{tilded couplings}).
The restriction $v_{\alpha\beta}=v$ when $\alpha\cap\beta=0$, while
otherwise it is zero, means that $\tilde M$ is a matrix with the
``two-packet'' RSB \cite{BrMo79,Cirano}, whose block diagonalization
is worked out in Appendix \ref{App5} for the most generic case.
In the scaling limit ($q\to 0$ and $v\sim q$), $\tilde M$ reduces
to the first two terms in (\ref{tilded couplings}),
and only $\bar W_1$ is kept,
as it is the only cubic coupling which is finite at criticality.
With the notations of Appendix \ref{App5}, we then have
\begin{equation}\label{scaling tilde M}
\begin{gathered}
\tilde M_1^{(1)}=\tilde M_1^{(2)}={{\tilde M}_1}^{\prime\,(2)}=\bar M_1,\\
\tilde M_2^{(1)}={{\tilde M}_2}^{\prime\,(1)}=\tilde M_2^{(2)}=
{{\tilde M}_2}^{\prime\,(2)}=\bar M_2,\\
\tilde M_2^{(3)}={{\tilde M}_2}^{\prime\,(3)}=\bar M_2-\bar W_1\,v,\\
\tilde M_3^{(1)}=\tilde M_3^{(2)}={{\tilde M}_3}^{\prime\,(2)}=
\tilde M_3^{(3)}={{\tilde M}_3}^{\prime\,(3)}=\tilde M_3^{(4)}=
\bar M_3.
\end{gathered}
\end{equation}
As it turns out from Appendix \ref{App5}, the two replicon classes are
independent of $v$ in this approximation, and they do not contribute to
$g_1$. To simplify the following algebra, and exploiting the arbitrariness
of $p$, the symmetric choice $p=n-p=n/2$ in computing the longitudinal (L)
and anomalous (A) eigenvalues will be applied.
After substituting the mass elements from
Eqs.\ (\ref{scaling tilde M}) and (\ref{bare masses}) into the L and A
blocks of Appendix \ref{App5}, we get
\begin{equation}\label{eigenvalues}
\begin{aligned}
\lambda^0_{\text{L}}&=2\bar w_1q\\[5pt]
\lambda^{\pm}_{\text{L}}&=\bar w_1q\,\Big\{1\pm\Big[1+n(n-2)\Big(1+\frac{v}{q}\Big
)^2
\Big]^{\frac{1}{2}}\Big\}\\[5pt]
\lambda^{\pm}_{\text{A}}&=\bar w_1q\,\Big\{\frac{n+2}{2}
\pm\Big[1+\frac{1}{4}n(n-4)
(1+\frac{v}{q}\Big
)^2
\Big]^{\frac{1}{2}}\Big\}
\end{aligned}\qquad\qquad\quad
p=\frac{n}{2},\quad q\to 0\quad \text{and} \quad v\sim q.
\end{equation}
Putting these expressions into (\ref{g1}), with the multiplicities taken from
Appendix \ref{App5}, the following result is obtained for the one-loop
scaling contribution:
\begin{equation}\label{g1 result}
g_1(v)=4\,\frac{n-2}{n}\,q\Big(1+\frac{v}{q}\Big)\,{\bar w}_1^2\,\frac{1}{N}
\sum_{\mathbf p}\bigg[\frac{1}{(p^2+\lambda^+_{\text{L}})
(p^2+\lambda^-_{\text{L}})}+\frac{n-4}{4}
\frac{1}{(p^2+\lambda^+_{\text{A}}) (p^2+\lambda^-_{\text{A}})}\bigg].
\end{equation}
(For $d<6$, ${\bar w}_1$ must be replaced by $w_1^*$.) It is obvious from
(\ref{eigenvalues}) and (\ref{g1 result}) that $g_1$ satisfies
\[
g_1(-v-2q)=-g_1(v).
\]
By a direct calculation of the second and third terms of (\ref{basic}), it was
checked that $g_1(v)$ in (\ref{g1 result}) ensures that the generic property
$g(v=0)=g'(v=0)=0$ is valid up to one-loop order.

By straightforward, though somewhat lengthy, manipulation of Eqs.\ %
(\ref{basic}) and (\ref{g1 result}) --- whilst using
the definitions of the $v$-independent masses in (\ref{free propagators})
and the $v$-dependent ones in (\ref{eigenvalues}) ---, the following
convenient form for the one-loop leading scaling term of $g(v)$ can
be concluded:
\begin{multline}\label{g result}
g(v)=\bar{g}(v)+4(n-2)\, \bar{w}_1\, (\bar{w}_1q)^3\,
(v/q)^2\,(3+v/q)\,\,\frac{1}{N}\sum_{\mathbf{p}}
\bigg[\frac{n-2}{(p^2+\lambda^+_{\text{L}})(p^2+\lambda^-_{\text{L}})
(p^2+r_{\text{L}})(p^2+r_{\text{R}})}\\[8pt]+
\frac{(n-4)^2/16}{(p^2+\lambda^+_{\text{A}})(p^2+\lambda^-_{\text{A}})
(p^2+r_{\text{A}})(p^2+r_{\text{R}})}\bigg]
\,\,+\,\,8n(n-2)\, \bar{w}_1\, (\bar{w}_1q)^5\,
(v/q)^2\,(2+v/q)\\[8pt]
\times\,
\frac{1}{N}\sum_{\mathbf{p}}
\bigg[\frac{(n-2)^2}{(p^2+\lambda^+_{\text{L}})(p^2+\lambda^-_{\text{L}})
(p^2+r_{\text{L}})^2(p^2+r_{\text{R}})^2}+
\frac{(n-4)^3/64}{(p^2+\lambda^+_{\text{A}})(p^2+\lambda^-_{\text{A}})
(p^2+r_{\text{A}})^2(p^2+r_{\text{R}})^2}\bigg]\,.
\end{multline}
This formula was derived for $p=n/2$ in the scaling limit
$q\to 0$ and $v\sim q$. In what follows, Eq.\ (\ref{g result}) is used to
track the behaviour of $g(v)$, and along with it the question of stability
of the RS spin glass phase, starting from high dimension ($d>8$) and ending
below the upper critical dimension ($d\lesssim
6$), where the nontrivial fixed point
is small, and one-loop results can be translated to the leading behaviour
of $g(v)$ in an $\epsilon=6-d$ expansion.

\subsection{The mean field like high dimensional regime: $d>8$}

In this case, the bare replicon generator $\bar{g}(v)$ as well possesses
the properties of Eq.\ (\ref{generic k}) as the exact one. From this and the
definition in (\ref{basic}), the leading term for $q\to 0$ and $q\sim v$
follows as
\begin{equation}\label{gbar}
\bar{g}(v)=\frac{1}{6}\,\big[2\bar{u}_2+n\bar{u}_3+\frac{1}{2}n^2(\bar{u}_1+2
\bar{u}_4)\big]\,q^3\,(v/q)^2\,(3+v/q),
\end{equation}
where (\ref{U0}) of Appendix \ref{App3} has been used for $p=n/2$,
and only the four quartic couplings remaining finite at $T_c$ are kept.
[To reproduce the mean field results of the Sherrington-Kirkpatrick model
\cite{SK}, we must substitute the values $\bar{u}_1=3$, $\bar{u}_2=2$,
$\bar{u}_3=-6$ and $\bar{u}_4=0$ into (\ref{gbar}).
We must be careful, however, if a direct comparison with Eq.\
(\ref{ghat}) is to be made, as it is valid in the limit $p\to 0$,
whereas $p=n/2$ was chosen in this section. Obviously, disregarding the
$n^2$ term in (\ref{gbar}), which comes from $p(n-p)$, leads to complete
agreement.]
Putting together this result and the high momentum leading term of
(\ref{g result}), $g(v)$ assumes the scaling form $q^3\,\hat g(v/q)$,
with
\[
\hat g(x)=
\bigg\{\frac{1}{6}\,\big[2\bar{u}_2+n\bar{u}_3+\frac{1}{2}n^2(\bar{u}_1+2
\bar{u}_4)\big]+
\frac{1}{4}\,(n-2)(n^2+8n-16)\,\,\bar{w}_1^4\,\,\frac{1}{N}\sum_{\mathbf p}
\frac{1}{p^8}\bigg\}\times x^2(3+x)
\]
A comparison with (\ref{ghat}) reveals the surprising fact that the
scaling function $\hat g(x)\sim x^2(3+x)$ remains very robust with respect
to mean field theory ($d=\infty$), apart from a normalization factor,
at least up to one-loop order.

The replicon mass for finite $n$ can now be easily derived from (\ref{k=0})
and (\ref{Ward3}):
\begin{equation}\label{2m1}
2m_1\cong n\,w_1 q-\frac{g(-2q)}{2q}.
\end{equation}
For $n>0$, there is a temperature domain below $T_c$ where the RS phase
is stable ($2m_1>0$). The condition $2m_1=0$ signals the end of the stable
RS phase, defining an Almeida-Thouless line in the $T$-$n$
\cite{Ko83}, or equivalently
in the $q$-$n$ plane (instead of the original AT line in the
temperature-magnetic field plane):
\begin{equation}\label{qAT1}
q_{\text{AT}}\cong C_d\,n,
\end{equation}
where the dimension-dependent slope has a finite limit for $d\to \infty$
(the zero-loop mean field limit), whereas it goes monotonically to zero
while $d\to 8$; see Fig.\ \ref{Fig}. A first order perturbative correction
leads to a shrinking RS phase (or growing RSB phase) when decreasing the
space dimension.

\subsection{The transitional regime: $6<d<8$}

Although critical behaviour is still governed by the Gaussian fixed point,
and standard perturbation expansion remains valid (no accumulating
infrared divergences), scaling of $g(v)$ is now modified to
\[
g(v)=\bar{w}_1\,(\bar{w}_1 q)^{2-\epsilon/2}\,\hat{g}(v/q),
\qquad \epsilon=6-d.
\]
The scaling function $\hat{g}$ follows from (\ref{g result}) after
sending the cutoff to infinity and rescaling the momentum $p^2
\rightarrow p^2/(\bar{w}_1 q)$. (The bare contribution $\bar{g}(v)$ is
subleading for $d<8$, and it can be neglected in what follows.) The following
properties are easily concluded:
\begin{enumerate}
\item
For $n=0$, the mean field form $\hat{g}\sim x^2(3+x)$  is again recovered,
together with its immediate aftermaths in Eqs.\ (\ref{diff g})
and (\ref{m1 vs w2}). The truncated form of (\ref{truncated g}) then follows
as well. This is, however, not at all trivial when $d<8$ --- as it was pointed
out in \cite{droplet} ---, since all the terms in (\ref{g}) are now the same
order, and a cancellation of the terms higher than cubic order in $v$ must
occur. This solves the problem raised in Ref.\ \cite{droplet}: the infinite
sum of its Eq.\ (11) is exactly zero, leading to instability. This conclusion
is now reached by the evaluation of the $g(v)$ function, and the previous
statements are valid, of course, in this first order perturbative framework.
\item
A nonanalytic $n$-dependence develops for $d<8$, which is proportional
to $n^{1-\epsilon/2}$. Although this nonanalycity is innocuous with respect
to the zero $n$ limit, the truncated form (\ref{truncated g}) is no longer
valid for finite $n$.
\end{enumerate}

The boundary of the RS state follows again from (\ref{2m1})
and (\ref{g result}), the zero $n$ limit of the propagators taken from
(\ref{free propagators}) and (\ref{eigenvalues}). The AT line in the
$q$-$n$ plane is now nonanalytic, and the tendency of a shrinking RS phase
persists when approaching the upper critical dimension $6$
(see Fig.\ \ref{Fig}):
\begin{equation}\label{qAT2}
n\cong 16\,\bar{w}_1^2\,\frac{1}{N}\sum_{\mathbf p}\frac{1}{p^4(p^2+2)^2}
\,\,(\bar{w}_1q_{\text{AT}})^{|\epsilon|/2},\qquad -2<\epsilon=6-d<0.
\end{equation}

\subsection{Below the upper critical dimension: $d\lesssim 6$}

The nontrivial fixed point \cite{HaLuCh76}
$\bar{w}^{*2}_1=-\epsilon/(n-2)+\dots$ governs
critical behaviour below six dimensions. A scaling analysis in Ref.\
\cite{droplet} showed that $G^{(k+1)}\,q^{k-1}\sim \bar{w}^{*2}_1
\,(\bar{w}^{*}_1q)^{\gamma/\beta}$, yielding by Eq.\ (\ref{g}) in the
scaling limit
\begin{equation}\label{g d<6}
g(v)=\bar{w}^{*}_1\,(\bar{w}^{*}_1q)^{\gamma/\beta+1}\,\hat{g}(v/q),
\qquad d<6.
\end{equation}
With this definition above, $\hat g$ has a finite limit for $\epsilon\to
0$, and it obviously depends on $d$ and $n$.
Renormalization group calculations provide $\gamma/\beta+1=2+
\epsilon/2+\dots$, see \cite{HaLuCh76,Iveta}. Our result is consistent
with this form at the leading order in $\epsilon$, the scaling function
is the same as that of the transitional regime in the previous subsection,
except that the momentum integrals must be performed at $d=6$.
The first property of the previous subsection, concerning the $n=0$
behaviour of the scaling function, can be entirely taken over, whereas
the $n$-dependence of $\hat{g}$, which is harmless around $d=6$,
will be discussed in the conclusions.

Both terms in Eq.\ (\ref{2m1}) are now proportional to
$(\bar{w}^{*}_1q)^{\gamma/\beta}$, and we get, in the limit $n\to 0$,
by means of (\ref{g result}) the replicon mass
\begin{equation}\label{2m1 d<6}
2m_1\cong (\bar{w}^{*}_1q)^{\gamma/\beta}\,
\bigg[ n-8\,\epsilon\int\frac{d^6 p}{(2\pi)^6}\frac{1}{p^4(p^2+2)^2}
+O(\epsilon^2)\bigg],
\qquad \epsilon=6-d\gtrsim 0\quad \text{and}\quad n\gtrsim 0.
\end{equation}
We can see from this formula that the stable RS spin glass phase for
$n$ small but finite --- which does exist in mean field theory and even
for $d>6$ --- completely disappears below six dimensions, continuing in this
way the trend of a shrinking RS phase (see Fig.\ \ref{Fig}).

\section{Summary and conclusions}\label{Conclusion}

The Ising spin glass in zero magnetic field has been studied in this
paper, using the field theoretic representation of the replicated
model and concentrating on the question of stability of the replica
symmetric glassy phase. Any preconception concerning the RS spin glass
phase has been avoided, except that the Legendre-transformed free energy
preserves the symmetry of the replicated paramagnet, Eqs.\ (\ref{O})
and (\ref{invariance}), just as in the case of common phase transitions
with spontanous symmetry breaking. As a convinient way to study stability,
the replicon generator function $g(v)$ was introduced. On the one hand,
it was exactly computed for the Sherrington-Kirkpatrick model, whereas droplet
theory ideas were applied to guess its behaviour in case of a droplet picture
scenario. In this later case, a highly singular behaviour in the spin
glass limit ($n\to 0$) was found. We followed the fate of the mean field
($d=\infty$) scenario by a first order perturbative calculation down below
the upper critical dimension $6$, where the fixed point is small and
the perturbative renormalization group is valid. The following conclusions
have been reached:
\begin{itemize}
\item
The scaling limit of $g(v)$, $q\to 0$ and $v\sim q$, has the common form
for $n=0$:
\begin{equation}\label{scaling again}
g(v)\sim q^{\lambda}\,\hat g(v/q),\qquad \text{with} \qquad \lambda=
\begin{cases}
3,&\text{$d>8$ and in mean field,}\\
2-\epsilon/2,&\text{$6<d<8$,}\\
1+\gamma/\beta=2+\epsilon/2+O(\epsilon^2),&\text{$d\lesssim 6$;}
\end{cases}
\end{equation}
$\epsilon=6-d$. The scaling function has proved to be very robust, as it is
\begin{equation}\label{ghat again}
\hat g(x)\sim x^2(3+x)
\end{equation}
for all three cases.

This result is not unexpected for $d>8$, as it must be
true for the full $g(v)$ (even for any finite $n$) built up from the
{\em exact\/} vertices. This can be proven by applying a simple scaling analysis
around the Gaussian fixed point. Using (\ref{g}), (\ref{W0}) and
(\ref{U0}), the exact $g(v)$ can be separated into two parts:
\[
g(v)=\Big(w_2\,v^2+\frac{1}{3}u_2\,v^3+\dots\Big)+\sum_{k>\frac{d}{2}-1}
\frac{1}{k!}\,G^{(k+1)}v^k,\qquad d>8, \quad n=0.
\]
The first term represents the analytical part, the ellipsis dots
stand for its subdominant contributions with $3<k\leq \frac{d}{2}-1$,
whereas the second one is nonanalytical with all of its subterms 
being the same order for $q\to 0$ and $v\sim q$:
\begin{equation}\label{hhat}
\sum_{k>\frac{d}{2}-1}\frac{1}{k!}\,G^{(k+1)}v^k\cong
\sum_{k>\frac{d}{2}-1}c_k\,q^{2-\frac{\epsilon}{2}}\,(v/q)^k
\equiv q^{2-\frac{\epsilon}{2}}\,\hat{h}(v/q).
\end{equation}
The analytical part obviously dominates for $d>8$, and the mean field like
truncated form in Eq.\ (\ref{truncated g}) is valid for
the exact $g(v)$ (and even for a generic $n$ too).

In the transitional regime, $6<d<8$, the analytical part disappears, and
$g(v)$ takes the form as in (\ref{hhat}). There is now no reason to think
that the exact scaling function $\hat{h}$ has the mean field functional
form $x^2(3+x)$, since all the higher order vertices equally contribute.
Nevertheless, the first order calculation provides --- somewhat
surprisingly --- Eqs.\ (\ref{scaling again}) and (\ref{ghat again}),
suggesting that a cancellation of terms with vertices other than the cubic
and quartic ones must occur in one-loop order. A similar remark is valid
for the regime below the upper critical dimension too. It must be realized,
however, that the truncated form (\ref{truncated g}) or (\ref{ghat again})
holds only for the spin glass limit $n=0$, as opposed to the $d>8$ regime,
and an infinite number of terms contribute for $n$ finite.

\item
The mean field scenario is very robust for decreasing $d$, see Eqs.\
(\ref{scaling again}) and (\ref{ghat again}), leading to an unstable RS
phase for $n=0$ even below $6$ dimensions. By computing the boundary of
stability, $2m_1=0$, in finite $n$ (a kind of AT line in zero magnetic field),
even a shrinking RS phase has been found with respect to the mean field
picture (Fig.\ 
\ref{Fig}); at least in this first order perturbative approximation.
\begin{figure}[h]
\begin{pspicture}(11,7)
\psline[linewidth=0.5pt]{->}(1,1)(10,1)\psline[linewidth=0.5pt]{->}(1,1)(1,6)
\uput[dr](10,1){$\bar{w}_1q$}\uput[l](1,6){$n$}
\psline(1,1)(9,2.5)\psline(1,1)(8.8,4)
\rput*(8,1.6){RSB}\uput{8pt}[30](9,2.5){(a) mean field, $d=\infty$}
\uput{8pt}[20](8.8,4){(b) $d>8$}
\uput[r](3,3.5){(c) $6<d<8$}\uput[r](5.2,5.7){(d) $d\lesssim 6$}
\pscurve(1,1)(1.05,1.4)(1.2,2)(2.5,3.5)(8.5,4.8)
\pscurve(1,3.1)(2,4.3)(3,4.8)(8.5,5.5)
\uput[l](1,3.1){$n_c$}
\rput*(2,5.4){RS}
\end{pspicture}
\caption{\label{Fig}
Phase boundary of the RS-RSB transition (Almeida-Thouless line)
in the $q-n$ plane. This one-loop result shows a shrinking RS phase with
decreasing $d$. (a) and (b) represent (\ref{qAT1}), while the nonanalytic
feature of the AT line in the transitional regime ($6<d<8$), Eq.\
(\ref{qAT2}), is displayed in (c). The stable RS domain for small but finite
$n$ disappears for $d<6$ and $n<n_c\sim
\epsilon=6-d$, as shown in (d).}
\end{figure}
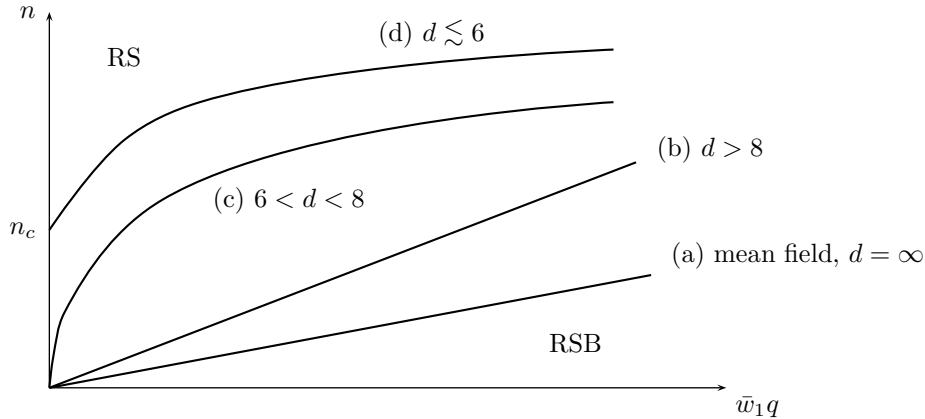

\item
The mean field formula Eq.\ (\ref{m1 vs w2}), which connects the replicon
mass ($2m_1$) and the replicon 3-point vertex ($w_2$) for $n=0$, remains
valid below 6 dimensions, at least in the leading $\epsilon$ order. This
follows from the robustness of $\hat g$, as shown in (\ref{scaling again})
and (\ref{ghat again}). From the independent calculation of the one-loop
3-point vertices \cite{Iveta} we can quote the relatively simple result
for $w_2$ (keeping $n$ finite here):
\begin{equation}\label{w2}
w_2=\frac{12}{(n-2)^2}\, w_1^*\,(w_1^*q)^{\gamma/\beta-1}
\int\frac{d^6 p}{(2\pi)^6}\,
\bigg[\frac{4-n}{(p^2+n)^2(p^2+2)^2}-\frac{n}{(p^2+n)^3(p^2+2)}\bigg]
\times \epsilon +O(\epsilon^2).
\end{equation}
Eq.\ (\ref{2m1 d<6}) follows immediately from (\ref{m1 vs w2}) in the spin glass
limit, providing us a good check of the theory.

\item
As it can be observed from the above discussion,
stability of the RS spin glass state in zero magnetic field, and thus the
possibility of the droplet scenario itself, can not occur in those high
dimensions where direct perturbation expansion (loop expansion) or the
perturbative renormalization group is applicable ($d\lesssim 6$). There is
even an obvious tendency that RSB is escalating with decreasing space
dimension. Although common objections against this statement might argue that
a droplet-like RS state is intrinsically nonperturbative, it must be stressed
that, at least for $d>8$, exact considerations are in accordance with the
(first order) perturbative result.

\item
The possibility for the emergence of a marginally stable droplet phase ---
notwithstanding the ever increasing RSB domain --- in low enough dimensions
will be put forward in this 
remark. The $n$-dependent
scaling form of $g(v)$ for the droplet picture, Eq.\ (\ref{n scaling}),
proposed in Section \ref{Droplet} clearly shows that $\hat g$ for the
droplet model should be a nontrivial generalized homogeneous function of
$v/q$ and $n$ near $T_c$. Just below 6 dimensions, $\hat g$ of (\ref{g d<6})
is not such, as it can be decomposed into the form
$\hat g=\hat{g}_{\text{an}}+\hat{g}_{\text{nonan}}$, where $\hat{g}_{\text{an}}$
is analytical ($\hat{g}_{\text{nonan}}$ is nonanalytical) for $n\to 0$,
respectively, and $\hat{g}_{\text{nonan}}\ll\hat{g}_{\text{an}}$
for any $v/q$. ($\hat{g}_{\text{nonan}}\sim n^{1-\epsilon/2}$ when
$6<d<8$.) As a matter of fact, $n$ in $\hat g$ plays the role of a small
mass, the ``replicon'' mass, as contrasted to the large mass, the
``longitudinal'' one. This small mass is, however, innocuous around $d=6$,
since it does not cause infrared problems. (See as an example Eq.\
(\ref{w2}): an infrared divergence would be expected at $d=4$,
if the collapse of the perturbative approach were neglected.)
We can postulate a dimension $d_\text{R}$, which is only naively believed to be
$4$, where $\hat{g}_{\text{nonan}}$ stops being negligible, and $\hat g$
may become a nontrivial generalized homogeneous function of $v/q$ and $n$.
Although a proof of this scenario and an estimate of $d_\text{R}$ seem to be
a rather hard task --- due to the nonperturbative nature of the problem
in these low dimensions ---,
these considerations could help the understanding
how a marginally stable replica symmetric phase
may emerge in low enough dimensions.\footnote{$d_\text{R}$, the lower
critical dimension of RSB, should not be missed with $d_l$, the lower
critical dimension of the spin glass state. Of course, the transition from
RSB to the droplet scenario can occur only if $d_l<d_\text{R}$.}
\end{itemize}

Our final comment concerns the work of Bray and Moore \cite{BrMo79}: In this
paper the authors use the same ``two-packet'' RSB for temporarily breaking
the replica symmetry, and eventually restoring the RS state by sending
$p$ (which they call $m$) to infinity. What is more, this RS state is
marginally stable! It must be stressed, however, that the two approaches
differ substantially: $p<n$ is always assumed here, and the ``two-packet''
RSB calculation used in this paper is only a tool to compute the replicon
generator of the RS state. The marginally stable RS phase found in
Ref.\ \cite{BrMo79} is obviously in conflict with our results, since the
replica symmetric spin glass state has proved to be unstable in the
regime $6<d<8$, assuming only the invariance of the Legendre-transformed
free energy, Eq.\ (\ref{invariance}), as a prerequisite.

\appendix
\section{Basic definitions}\label{App1}

In this appendix the definitions of the lower and upper case vertices are
displayed in a tabular form for the generic RS state. Although not all of
them are used in this paper, completeness was favoured up to the quartic order.
Due to the RS symmetry, the number of different vertices is limited to
3 for the masses ($k=2$), 8 for the cubic ($k=3$) and 23 for the quartic
($k=4$) ones. The first column provides an obvious graphical representation,
while in the second one the lower case bare couplings are enumerated
together with their corresponding invariants; see Eqs.\ (\ref{L}) and
(\ref{LI}). The alternative set of vertices with the upper case notation
is best defined by the derivatives of the {\em exact\/}
Legendre-transformed free
energy $F$, Eq.\ (\ref{vertices}), and they are listed in the last column.
The corresponding {\em bare\/} couplings enter the Lagrangian in the following
way (see footnote \ref{fn2}):
\begin{equation}\label{bareM}
\mathcal{L}=\frac{1}{2}\sum_{\mathbf p}\,\sum_{(\alpha\beta),(\gamma\delta)}
\big(p^2\,\delta^{\text{Kr}}_{\alpha\beta,\gamma\delta}+
\bar{M}_{\alpha\beta,\gamma\delta}\big)\,\phi^{\alpha\beta}_{\mathbf p}
\phi^{\gamma\delta}_{-{\mathbf p}}+\mathcal{L}^{\text{I}},
\end{equation}
with the interaction part
\begin{equation}\label{bareG}
\mathcal{L}^{\text{I}}=-N^{\frac{1}{2}}\,\bar{H}\,\sum_{(\alpha\beta)}
\,\phi^{\alpha\beta}_{{\mathbf p}=0}
-\sum_{k=3}
\,\frac{1}{k!\,N^{\frac{k}{2}-1}}\,
\sideset{}{'}\sum_{{\mathbf {p}}_i}\,
\sum_{(\alpha_1\beta_1),\dots ,(\alpha_k\beta_k)}
\bar{G}^{(k)}_{\alpha_1\beta_1,\dots ,\alpha_k\beta_k}\,
\phi^{\alpha_1\beta_1}_{{\mathbf p}_1}\ldots\,
\phi^{\alpha_k\beta_k}_{{\mathbf p}_k}.
\end{equation}
[$\bar{H}=2\bar{h}$ follows trivially
by comparing (\ref{LI}) and (\ref{bareG}).]

As already declared in the main text, the common notations are used in what
follows for the quadratic ($-\bar m_j$ and $-M_j$), cubic ($\bar w_j$ and $W_j$)
and quartic
($\bar u_j$ and $U_j$) vertices (instead of $\bar g^{(k)}_j$ and
$G^{(k)}_j$, $k=2,3$ and $4$).

\begin{itemize}
\item
The masses, $k=2$:
\begin{center}
\begin{tabular}{|c|c|c|c|}\hline
$\quad j\quad$&&$\bar{g}^{(k)}_jI^{(k)}_j$&$G^{(k)}_j$\\\hline\hline
$\quad 1\quad$&\begin{pspicture}[0.52](2,1)
\rput(0.5,0.5){\dotnode{A}{}}
\rput(1.5,0.5){\dotnode{B}{}}
\nccurve[angleA=35,angleB=145]{A}{B}
\nccurve[angleA=-35,angleB=-145]{A}{B}
\end{pspicture}
&
$\displaystyle -\bar{m}_1\,\sum_{\alpha\beta}
\phi^{\alpha\beta}_{{\mathbf p}_1}\phi^{\alpha\beta}_{{\mathbf p}_2}$
&$-M_1=-M_{\alpha\beta,\alpha\beta}$\\\hline
$\quad 2\quad$&\begin{pspicture}[0.52](2,1)
\rput(0.4,0.7){\dotnode{A}{}}
\rput(1.6,0.7){\dotnode{B}{}}
\rput(1,0.4){\dotnode{C}{}}
\ncline{A}{C}\ncline{C}{B}
\end{pspicture}
&
$\displaystyle -\bar{m}_2\,\sum_{\alpha\beta\gamma}
\phi^{\alpha\gamma}_{{\mathbf p}_1}\phi^{\beta\gamma}_{{\mathbf p}_2}$
&$-M_2=-M_{\alpha\gamma,\beta\gamma}$\\\hline
$\quad 3\quad$&\begin{pspicture}[0.52](2,1)
\rput(0.5,0.3){\dotnode{A}{}}
\rput(1.5,0.3){\dotnode{B}{}}
\rput(0.5,0.7){\dotnode{C}{}}
\rput(1.5,0.7){\dotnode{D}{}}
\ncline{A}{B}\ncline{C}{D}
\end{pspicture}
&
$\displaystyle -\bar{m}_3\,\sum_{\alpha\beta\gamma\delta}
\phi^{\alpha\beta}_{{\mathbf p}_1}\phi^{\gamma\delta}_{{\mathbf p}_2}$
&$-M_3=-M_{\alpha\beta,\gamma\delta}$\\\hline
\end{tabular}
\end{center}

\item
Cubic vertices, $k=3$:
\begin{center}
\begin{tabular}{|c|c|c|c|}\hline
$\quad j\quad$&&$\bar{g}^{(k)}_jI^{(k)}_j$&$G^{(k)}_j$\\\hline\hline
$\quad 1 \quad$ &
\begin{pspicture}[0.52](2,1)
\rput(0.55,0.2){\dotnode{A}{}}\rput(1.45,0.2){\dotnode{B}{}}
\rput(1,0.8){\dotnode{C}{}}
\ncline{A}{B}\ncline{B}{C}\ncline{C}{A}
\end{pspicture} &
$\displaystyle \bar{w}_1  \,\sum_{\alpha\beta\gamma}
\phi^{\alpha\beta}_{{\mathbf p}_1}\phi^{\beta\gamma}_{{\mathbf p}_2}
\phi^{\gamma\alpha}_{{\mathbf p}_3}$
&$W_1=W_{\alpha\beta,\beta\gamma,\gamma\alpha}$\\\hline
$\quad 2 \quad$ &
\begin{pspicture}[0.52](2,1)
\rput(0.5,0.5){\dotnode{A}{}}
\rput(1.5,0.5){\dotnode{B}{}}
\nccurve[angleA=45,angleB=135]{A}{B}
\nccurve[angleA=-45,angleB=-135]{A}{B}
\ncline{A}{B}
\end{pspicture} &
$\displaystyle \bar{w}_2  \,\sum_{\alpha\beta}
\phi^{\alpha\beta}_{{\mathbf p}_1}
\phi^{\alpha\beta}_{{\mathbf p}_2}\phi^{\alpha\beta}_{{\mathbf p}_3}$
&$W_2=W_{\alpha\beta,\alpha\beta,\alpha\beta}$\\\hline
$\quad 3 \quad$ &
\begin{pspicture}[0.52](2,1)
\rput(0.2,0.5){\dotnode{A}{}}\rput(1,0.5){\dotnode{B}{}}
\rput(1.8,0.5){\dotnode{C}{}}
\nccurve[angleA=35,angleB=145]{A}{B}
\nccurve[angleA=-35,angleB=-145]{A}{B}
\ncline{B}{C}
\end{pspicture} &
$\displaystyle \bar{w}_3  \,\sum_{\alpha\beta\gamma}
\phi^{\alpha\beta}_{{\mathbf p}_1}
\phi^{\alpha\beta}_{{\mathbf p}_2}\phi^{\beta\gamma}_{{\mathbf p}_3}$
&$W_3=W_{\alpha\beta,\alpha\beta,\beta\gamma}$\\\hline
$\quad 4 \quad$ &
\begin{pspicture}[0.52](2,1)
\rput(0.2,0.5){\dotnode{A}{}}\rput(0.8,0.5){\dotnode{B}{}}
\rput(1.2,0.5){\dotnode{C}{}}\rput(1.8,0.5){\dotnode{D}{}}
\nccurve[angleA=35,angleB=145]{A}{B}
\nccurve[angleA=-35,angleB=-145]{A}{B}
\ncline{C}{D}
\end{pspicture} &
$\displaystyle \bar{w}_4  \,\sum_{\alpha\beta\gamma\delta}
\phi^{\alpha\beta}_{{\mathbf p}_1}
\phi^{\alpha\beta}_{{\mathbf p}_2}\phi^{\gamma\delta}_{{\mathbf p}_3}$
&$W_4=W_{\alpha\beta,\alpha\beta,\gamma\delta}$\\\hline
$\quad 5 \quad$ &
\begin{pspicture}[0.52](2,1)
\rput(0.7,0.25){\dotnode{A}{}}\rput(1.3,0.25){\dotnode{B}{}}
\rput(0.35,0.65){\dotnode{C}{}}\rput(1.65,0.65){\dotnode{D}{}}
\ncline{A}{B}\ncline{A}{C}\ncline{B}{D}
\end{pspicture} &
$\displaystyle \bar{w}_5  \,\sum_{\alpha\beta\gamma\delta}
\phi^{\alpha\beta}_{{\mathbf p}_1}
\phi^{\alpha\gamma}_{{\mathbf p}_2}\phi^{\beta\delta}_{{\mathbf p}_3}$
&$W_5=W_{\alpha\beta,\alpha\gamma,\beta\delta}$\\\hline
$\quad 6 \quad$ &
\begin{pspicture}[0.52](2,1)
\rput(1,0.5){\dotnode{A}{}}\rput(1,0.85){\dotnode{B}{}}
\rput(0.7,0.2){\dotnode{C}{}}\rput(1.3,0.2){\dotnode{D}{}}
\ncline{A}{B}\ncline{A}{C}\ncline{A}{D}
\end{pspicture} &
$\displaystyle \bar{w}_6  \,\sum_{\alpha\beta\gamma\delta}
\phi^{\alpha\beta}_{{\mathbf p}_1}
\phi^{\alpha\gamma}_{{\mathbf p}_2}\phi^{\alpha\beta}_{{\mathbf p}_3}$
&$W_6=W_{\alpha\beta,\alpha\gamma,\alpha\delta}$\\\hline
$\quad 7 \quad$ &
\begin{pspicture}[0.52](2,1)
\rput(1,0.55){\dotnode{C}{}}\rput(0.6,0.75){\dotnode{A}{}}
\rput(1.4,0.75){\dotnode{B}{}}
\rput(0.6,0.3){\dotnode{D}{}}\rput(1.4,0.3){\dotnode{M}{}}
\ncline{C}{A}\ncline{C}{B}\ncline{D}{M}
\end{pspicture} &
$\displaystyle \bar{w}_7 \,\sum_{\alpha\beta\gamma\delta\mu}
\phi^{\alpha\gamma}_{{\mathbf p}_1}\phi^{\beta\gamma}_{{\mathbf p}_2}
\phi^{\delta\mu}_{{\mathbf p}_3}$
&$W_7=W_{\alpha\gamma,\beta\gamma,\delta\mu}$\\\hline
$\quad 8 \quad$ &
\begin{pspicture}[0.52](2,1)
\rput(0.6,0.2){\dotnode{A}{}}\rput(1.4,0.2){\dotnode{B}{}}
\rput(0.6,0.5){\dotnode{C}{}}\rput(1.4,0.5){\dotnode{D}{}}
\rput(0.6,0.8){\dotnode{M}{}}\rput(1.4,0.8){\dotnode{N}{}}
\ncline{A}{B}\ncline{C}{D}\ncline{M}{N}
\end{pspicture} &
$\displaystyle \bar{w}_8 \,\sum_{\alpha\beta\gamma\delta\mu\nu}
\phi^{\alpha\beta}_{{\mathbf p}_1}\phi^{\gamma\delta}_{{\mathbf p}_2}
\phi^{\mu\nu}_{{\mathbf p}_3}$
&$W_8=W_{\alpha\beta,\gamma\delta,\mu\nu}$\\\hline
\end{tabular}
\end{center}

\item Quartic vertices, $k=4$:
\begin{center}
\begin{tabular}{|c|c|c|c|}\hline
$\quad j\quad$&&$\bar{g}^{(k)}_jI^{(k)}_j$&$G^{(k)}_j$\\\hline\hline
$\quad 1 \quad$ &
\begin{pspicture}[0.52](2,1)
\rput(0.6,0.2){\dotnode{A}{}}\rput(1.4,0.2){\dotnode{B}{}}
\rput(0.6,0.8){\dotnode{D}{}}\rput(1.4,0.8){\dotnode{C}{}}
\ncline{A}{B}\ncline{B}{C}\ncline{C}{D}\ncline{D}{A}
\end{pspicture} &
$\displaystyle \bar{u}_1\,\sum_{\alpha\beta\gamma\delta}
\phi^{\alpha\beta}_{{\mathbf p}_1}\phi^{\beta\gamma}_{{\mathbf p}_2}
\phi^{\gamma\delta}_{{\mathbf p}_3}\phi^{\delta\alpha}_{{\mathbf p}_4}$
&$U_1=U_{\alpha\beta,\beta\gamma,\gamma\delta,\delta\alpha}$\\\hline
$\quad 2 \quad$ &
\begin{pspicture}[0.52](2,1)
\rput(0.5,0.5){\dotnode{A}{}}
\rput(1.5,0.5){\dotnode{B}{}}
\nccurve[angleA=25,angleB=155]{A}{B}
\nccurve[angleA=-25,angleB=-155]{A}{B}
\nccurve[angleA=65,angleB=115]{A}{B}
\nccurve[angleA=-65,angleB=-115]{A}{B}
\end{pspicture} &
$\displaystyle \bar{u}_2\,\sum_{\alpha\beta}
\phi^{\alpha\beta}_{{\mathbf p}_1}\phi^{\alpha\beta}_{{\mathbf p}_2}
\phi^{\alpha\beta}_{{\mathbf p}_3}\phi^{\alpha\beta}_{{\mathbf p}_4}$
&$U_2=U_{\alpha\beta,\alpha\beta,\alpha\beta,\alpha\beta}$\\\hline
$\quad 3 \quad$ &
\begin{pspicture}[0.52](2,1)
\rput(0.2,0.5){\dotnode{A}{}}\rput(1,0.5){\dotnode{C}{}}
\rput(1.8,0.5){\dotnode{B}{}}
\nccurve[angleA=35,angleB=145]{A}{C}
\nccurve[angleA=-35,angleB=-145]{A}{C}
\nccurve[angleA=35,angleB=145]{C}{B}
\nccurve[angleA=-35,angleB=-145]{C}{B}
\end{pspicture} &
$\displaystyle \bar{u}_3\,\sum_{\alpha\beta\gamma}
\phi^{\alpha\gamma}_{{\mathbf p}_1}\phi^{\alpha\gamma}_{{\mathbf p}_2}
\phi^{\beta\gamma}_{{\mathbf p}_3}\phi^{\beta\gamma}_{{\mathbf p}_4}$
&$U_3=U_{\alpha\gamma,\alpha\gamma,\beta\gamma,\beta\gamma}$\\\hline
$\quad 4 \quad$ &
\begin{pspicture}[0.52](2,1)
\rput(0.2,0.5){\dotnode{A}{}}\rput(0.8,0.5){\dotnode{B}{}}
\rput(1.2,0.5){\dotnode{C}{}}\rput(1.8,0.5){\dotnode{D}{}}
\nccurve[angleA=35,angleB=145]{A}{B}
\nccurve[angleA=-35,angleB=-145]{A}{B}
\nccurve[angleA=35,angleB=145]{C}{D}
\nccurve[angleA=-35,angleB=-145]{C}{D}
\end{pspicture} &
$\displaystyle \bar{u}_4\,\sum_{\alpha\beta\gamma\delta}
\phi^{\alpha\beta}_{{\mathbf p}_1}\phi^{\alpha\beta}_{{\mathbf p}_2}
\phi^{\gamma\delta}_{{\mathbf p}_3}\phi^{\gamma\delta}_{{\mathbf p}_4}$
&$U_4=U_{\alpha\beta,\alpha\beta,\gamma\delta,\gamma\delta}$\\\hline
$\quad 5 \quad$ &
\begin{pspicture}[0.52](2,1)
\rput(0.55,0.2){\dotnode{A}{}}\rput(1.45,0.2){\dotnode{B}{}}
\rput(1,0.8){\dotnode{C}{}}
\nccurve[angleA=30,angleB=150]{A}{B}
\nccurve[angleA=-30,angleB=-150]{A}{B}
\ncline{B}{C}\ncline{C}{A}
\end{pspicture} &
$\displaystyle \bar{u}_5\,\sum_{\alpha\beta\gamma}
\phi^{\alpha\beta}_{{\mathbf p}_1}\phi^{\alpha\beta}_{{\mathbf p}_2}
\phi^{\alpha\gamma}_{{\mathbf p}_3}\phi^{\beta\gamma}_{{\mathbf p}_4}$
&$U_5=U_{\alpha\beta,\alpha\beta,\alpha\gamma,\beta\gamma}$\\\hline
$\quad 6 \quad$ &
\begin{pspicture}[0.52](2,1)
\rput(0.5,0.25){\dotnode{A}{}}\rput(0.5,0.75){\dotnode{B}{}}
\rput(1,0.5){\dotnode{C}{}}\rput(1.5,0.5){\dotnode{D}{}}
\ncline{A}{B}\ncline{A}{C}\ncline{B}{C}\ncline{C}{D}
\end{pspicture} &
$\displaystyle \bar{u}_6\,\sum_{\alpha\beta\gamma\delta}
\phi^{\alpha\beta}_{{\mathbf p}_1}\phi^{\alpha\gamma}_{{\mathbf p}_2}
\phi^{\beta\gamma}_{{\mathbf p}_3}\phi^{\gamma\delta}_{{\mathbf p}_4}$
&$U_6=U_{\alpha\beta,\alpha\gamma,\beta\gamma,\gamma\delta}$\\\hline
$\quad 7 \quad$ &
\begin{pspicture}[0.52](2,1)
\rput(0.35,0.25){\dotnode{A}{}}\rput(0.35,0.75){\dotnode{B}{}}
\rput(0.85,0.5){\dotnode{C}{}}
\rput(1.15,0.5){\dotnode{D}{}}
\rput(1.65,0.5){\dotnode{M}{}}
\ncline{A}{B}\ncline{A}{C}\ncline{B}{C}
\ncline{D}{M}
\end{pspicture} &
$\displaystyle \bar{u}_7\,\sum_{\alpha\beta\gamma\delta\mu}
\phi^{\alpha\beta}_{{\mathbf p}_1}\phi^{\alpha\gamma}_{{\mathbf p}_2}
\phi^{\beta\gamma}_{{\mathbf p}_3}\phi^{\delta\mu}_{{\mathbf p}_4}$
&$U_7=U_{\alpha\beta,\alpha\gamma,\beta\gamma,\delta\mu}$\\\hline
$\quad 8 \quad$ &
\begin{pspicture}[0.52](2,1)
\rput(0.4,0.5){\dotnode{A}{}}\rput(1,0.5){\dotnode{B}{}}
\rput(1.6,0.5){\dotnode{C}{}}
\nccurve[angleA=45,angleB=135]{A}{B}
\nccurve[angleA=-45,angleB=-135]{A}{B}
\ncline{A}{B}\ncline{B}{C}
\end{pspicture} &
$\displaystyle \bar{u}_8\,\sum_{\alpha\beta\gamma}
\phi^{\alpha\beta}_{{\mathbf p}_1}\phi^{\alpha\beta}_{{\mathbf p}_2}
\phi^{\alpha\beta}_{{\mathbf p}_3}\phi^{\beta\gamma}_{{\mathbf p}_4}$
&$U_8=U_{\alpha\beta,\alpha\beta,\alpha\beta,\beta\gamma}$\\\hline
$\quad 9 \quad$ &
\begin{pspicture}[0.52](2,1)
\rput(0.25,0.5){\dotnode{A}{}}\rput(0.85,0.5){\dotnode{B}{}}
\rput(1.15,0.5){\dotnode{C}{}}\rput(1.75,0.5){\dotnode{D}{}}
\nccurve[angleA=45,angleB=135]{A}{B}
\nccurve[angleA=-45,angleB=-135]{A}{B}
\ncline{A}{B}\ncline{C}{D}
\end{pspicture} &
$\displaystyle \bar{u}_9\,\sum_{\alpha\beta\gamma\delta}
\phi^{\alpha\beta}_{{\mathbf p}_1}\phi^{\alpha\beta}_{{\mathbf p}_2}
\phi^{\alpha\beta}_{{\mathbf p}_3}\phi^{\gamma\delta}_{{\mathbf p}_4}$
&$U_9=U_{\alpha\beta,\alpha\beta,\alpha\beta,\gamma\delta}$\\\hline
$\quad 10 \quad$ &
\begin{pspicture}[0.52](2,1)
\rput(0.4,0.5){\dotnode{A}{}}\rput(1,0.5){\dotnode{B}{}}
\rput(1.4,0.25){\dotnode{C}{}}\rput(1.4,0.75){\dotnode{D}{}}
\nccurve[angleA=45,angleB=135]{A}{B}
\nccurve[angleA=-45,angleB=-135]{A}{B}
\ncline{B}{C}\ncline{B}{D}
\end{pspicture} &
$\displaystyle \bar{u}_{10}\,\sum_{\alpha\beta\gamma\delta}
\phi^{\alpha\beta}_{{\mathbf p}_1}\phi^{\alpha\beta}_{{\mathbf p}_2}
\phi^{\beta\gamma}_{{\mathbf p}_3}\phi^{\beta\delta}_{{\mathbf p}_4}$ 
&$U_{10}=U_{\alpha\beta,\alpha\beta,\beta\gamma,\beta\delta}$\\\hline
$\quad 11 \quad$ &
\begin{pspicture}[0.52](2,1)
\rput(0.25,0.5){\dotnode{A}{}}\rput(0.75,0.5){\dotnode{B}{}}
\rput(1.25,0.5){\dotnode{C}{}}\rput(1.75,0.5){\dotnode{D}{}}
\nccurve[angleA=45,angleB=135]{A}{B}
\nccurve[angleA=-45,angleB=-135]{A}{B}
\ncline{B}{C}\ncline{C}{D}
\end{pspicture} &
$\displaystyle \bar{u}_{11}\,\sum_{\alpha\beta\gamma\delta}
\phi^{\alpha\beta}_{{\mathbf p}_1}\phi^{\alpha\beta}_{{\mathbf p}_2}
\phi^{\beta\gamma}_{{\mathbf p}_3}\phi^{\gamma\delta}_{{\mathbf p}_4}$
&$U_{11}=U_{\alpha\beta,\alpha\beta,\beta\gamma,\gamma\delta}$\\\hline
$\quad 12 \quad$ &
\begin{pspicture}[0.52](2,1)
\rput(0.1,0.5){\dotnode{A}{}}\rput(0.6,0.5){\dotnode{B}{}}
\rput(1.1,0.5){\dotnode{C}{}}\rput(1.4,0.5){\dotnode{D}{}}
\rput(1.9,0.5){\dotnode{M}{}}
\nccurve[angleA=45,angleB=135]{A}{B}
\nccurve[angleA=-45,angleB=-135]{A}{B}
\ncline{B}{C}\ncline{D}{M}
\end{pspicture} &
$\displaystyle \bar{u}_{12}\,\sum_{\alpha\beta\gamma\delta\mu}
\phi^{\alpha\beta}_{{\mathbf p}_1}\phi^{\alpha\beta}_{{\mathbf p}_2}
\phi^{\beta\gamma}_{{\mathbf p}_3}\phi^{\delta\mu}_{{\mathbf p}_4}$
&$U_{12}=U_{\alpha\beta,\alpha\beta,\beta\gamma,\delta\mu}$\\\hline
$\quad 13 \quad$ &
\begin{pspicture}[0.52](2,1)
\rput(0.7,0.5){\dotnode{A}{}}\rput(1.3,0.5){\dotnode{B}{}}
\rput(0.3,0.75){\dotnode{C}{}}\rput(1.7,0.75){\dotnode{D}{}}
\nccurve[angleA=45,angleB=135]{A}{B}
\nccurve[angleA=-45,angleB=-135]{A}{B}
\ncline{A}{C}\ncline{B}{D}
\end{pspicture} &
$\displaystyle \bar{u}_{13}\,\sum_{\alpha\beta\gamma\delta}
\phi^{\alpha\beta}_{{\mathbf p}_1}\phi^{\alpha\beta}_{{\mathbf p}_2}
\phi^{\alpha\gamma}_{{\mathbf p}_3}\phi^{\beta\delta}_{{\mathbf p}_4}$
&$U_{13}=U_{\alpha\beta,\alpha\beta,\alpha\gamma,\beta\delta}$\\\hline
$\quad 14 \quad$ &
\begin{pspicture}[0.52](2,1)
\rput(0.25,0.5){\dotnode{A}{}}\rput(0.85,0.5){\dotnode{B}{}}
\rput(1.15,0.5){\dotnode{C}{}}
\rput(1.55,0.25){\dotnode{D}{}}\rput(1.55,0.75){\dotnode{M}{}}
\nccurve[angleA=45,angleB=135]{A}{B}
\nccurve[angleA=-45,angleB=-135]{A}{B}
\ncline{C}{D}\ncline{C}{M}
\end{pspicture} &
$\displaystyle \bar{u}_{14}\,\sum_{\alpha\beta\gamma\delta\mu}
\phi^{\alpha\beta}_{{\mathbf p}_1}\phi^{\alpha\beta}_{{\mathbf p}_2}
\phi^{\gamma\delta}_{{\mathbf p}_3}\phi^{\gamma\mu}_{{\mathbf p}_4}$
&$U_{14}=U_{\alpha\beta,\alpha\beta,\gamma\delta,\gamma\mu}$\\\hline
$\quad 15 \quad$ &
\begin{pspicture}[0.52](2,1)
\rput(0.25,0.5){\dotnode{A}{}}\rput(0.85,0.5){\dotnode{B}{}}
\rput(1.15,0.75){\dotnode{C}{}}\rput(1.55,0.75){\dotnode{D}{}}
\rput(1.15,0.25){\dotnode{M}{}}\rput(1.55,0.25){\dotnode{N}{}}
\nccurve[angleA=45,angleB=135]{A}{B}
\nccurve[angleA=-45,angleB=-135]{A}{B}
\ncline{C}{D}\ncline{M}{N}
\end{pspicture} &
$\displaystyle \bar{u}_{15}\,\sum_{\alpha\beta\gamma\delta\mu\nu}
\phi^{\alpha\beta}_{{\mathbf p}_1}\phi^{\alpha\beta}_{{\mathbf p}_2}
\phi^{\gamma\delta}_{{\mathbf p}_3}\phi^{\mu\nu}_{{\mathbf p}_4}$
&$U_{15}=U_{\alpha\beta,\alpha\beta,\gamma\delta,\mu\nu}$\\\hline
$\quad 16 \quad$ &
\begin{pspicture}[0.52](2,1)
\rput(0.6,0.3){\dotnode{B}{}}\rput(1.4,0.3){\dotnode{D}{}}
\rput(0.2,0.7){\dotnode{A}{}}\rput(1,0.7){\dotnode{C}{}}
\rput(1.8,0.7){\dotnode{M}{}}
\ncline{A}{B}\ncline{B}{C}\ncline{C}{D}\ncline{D}{M}
\end{pspicture} &
$\displaystyle \bar{u}_{16}\,\sum_{\alpha\beta\gamma\delta\mu}
\phi^{\alpha\beta}_{{\mathbf p}_1}\phi^{\beta\gamma}_{{\mathbf p}_2}
\phi^{\gamma\delta}_{{\mathbf p}_3}\phi^{\delta\mu}_{{\mathbf p}_4}$
&$U_{16}=U_{\alpha\beta,\beta\gamma,\gamma\delta,\delta\mu}$\\\hline
$\quad 17 \quad$ &
\begin{pspicture}[0.52](2,1)
\rput(0.25,0.25){\dotnode{A}{}}\rput(0.75,0.25){\dotnode{B}{}}
\rput(1.25,0.25){\dotnode{C}{}}\rput(1.75,0.25){\dotnode{D}{}}
\rput(0.75,0.75){\dotnode{M}{}}\rput(1.25,0.75){\dotnode{N}{}}
\ncline{A}{B}\ncline{B}{C}\ncline{C}{D}\ncline{M}{N}
\end{pspicture} &
$\displaystyle \bar{u}_{17}\,\sum_{\alpha\beta\gamma\delta\mu\nu}
\phi^{\alpha\beta}_{{\mathbf p}_1}\phi^{\beta\gamma}_{{\mathbf p}_2}
\phi^{\gamma\delta}_{{\mathbf p}_3}\phi^{\mu\nu}_{{\mathbf p}_4}$
&$U_{17}=U_{\alpha\beta,\beta\gamma,\gamma\delta,\mu\nu}$\\\hline
$\quad 18 \quad$ &
\begin{pspicture}[0.52](2,1)
\rput(0.35,0.5){\dotnode{A}{}}\rput(0.75,0.5){\dotnode{B}{}}
\rput(1.15,0.5){\dotnode{C}{}}
\rput(1.55,0.25){\dotnode{D}{}}\rput(1.55,0.75){\dotnode{M}{}}
\ncline{A}{B}\ncline{B}{C}\ncline{C}{D}\ncline{C}{M}
\end{pspicture} &
$\displaystyle \bar{u}_{18}\,\sum_{\alpha\beta\gamma\delta\mu}
\phi^{\alpha\beta}_{{\mathbf p}_1}\phi^{\beta\gamma}_{{\mathbf p}_2}
\phi^{\gamma\delta}_{{\mathbf p}_3}\phi^{\gamma\mu}_{{\mathbf p}_4}$
&$U_{18}=U_{\alpha\beta,\beta\gamma,\gamma\delta,\gamma\mu}$\\\hline
$\quad 19 \quad$ &
\begin{pspicture}[0.52](2,1)
\rput(1,0.5){\dotnode{M}{}}
\rput(1.3,0.8){\dotnode{A}{}}\rput(0.7,0.8){\dotnode{B}{}}
\rput(0.7,0.2){\dotnode{C}{}}\rput(1.3,0.2){\dotnode{D}{}}
\ncline{M}{A}\ncline{M}{B}\ncline{M}{C}\ncline{M}{D}
\end{pspicture} &
$\displaystyle \bar{u}_{19}\,\sum_{\alpha\beta\gamma\delta\mu}
\phi^{\mu\alpha}_{{\mathbf p}_1}\phi^{\mu\beta}_{{\mathbf p}_2}
\phi^{\mu\gamma}_{{\mathbf p}_3}\phi^{\mu\delta}_{{\mathbf p}_4}$
&$U_{19}=U_{\mu\alpha,\mu\beta,\mu\gamma,\mu\delta}$\\\hline
$\quad 20 \quad$ &
\begin{pspicture}[0.52](2,1)
\rput(0.2,0.5){\dotnode{A}{}}\rput(0.6,0.5){\dotnode{B}{}}
\rput(1.8,0.5){\dotnode{C}{}}\rput(1.2,0.5){\dotnode{D}{}}
\rput(1.6,0.75){\dotnode{M}{}}\rput(1.6,0.25){\dotnode{N}{}}
\ncline{A}{B}\ncline{C}{D}\ncline{D}{M}\ncline{D}{N}
\end{pspicture} &
$\displaystyle \bar{u}_{20}\,\sum_{\alpha\beta\gamma\delta\mu\nu}
\phi^{\alpha\beta}_{{\mathbf p}_1}\phi^{\gamma\delta}_{{\mathbf p}_2}
\phi^{\gamma\mu}_{{\mathbf p}_3}\phi^{\gamma\nu}_{{\mathbf p}_4}$
&$U_{20}=U_{\alpha\beta,\gamma\delta,\gamma\mu,\gamma\nu}$\\\hline
$\quad 21 \quad$ &
\begin{pspicture}[0.52](2,1)
\rput(0.8,0.5){\dotnode{A}{}}\rput(1.2,0.5){\dotnode{B}{}} 
\rput(0.4,0.25){\dotnode{C}{}}\rput(0.4,0.75){\dotnode{D}{}} 
\rput(1.6,0.25){\dotnode{M}{}}\rput(1.6,0.75){\dotnode{N}{}}
\ncline{A}{C}\ncline{A}{D}\ncline{B}{M}\ncline{B}{N}
\end{pspicture} &
$\displaystyle \bar{u}_{21}\,\sum_{\alpha\beta\gamma\delta\mu\nu}
\phi^{\alpha\gamma}_{{\mathbf p}_1}\phi^{\alpha\delta}_{{\mathbf p}_2}
\phi^{\beta\mu}_{{\mathbf p}_3}\phi^{\beta\nu}_{{\mathbf p}_4}$
&$U_{21}=U_{\alpha\gamma,\alpha\delta,\beta\mu,\beta\nu}$\\\hline
$\quad 22 \quad$ &
\begin{pspicture}[0.52](2,1)
\rput(0.8,0.5){\dotnode{C}{}}\rput(0.4,0.75){\dotnode{B}{}}
\rput(0.4,0.25){\dotnode{A}{}}
\rput(1.2,0.25){\dotnode{D}{}}\rput(1.6,0.25){\dotnode{M}{}}
\rput(1.2,0.75){\dotnode{N}{}}\rput(1.6,0.75){\dotnode{R}{}}
\ncline{A}{C}\ncline{B}{C}\ncline{D}{M}\ncline{N}{R}
\end{pspicture} &
$\displaystyle \bar{u}_{22}\,\sum_{\alpha\beta\gamma\delta\mu\nu\rho}
\phi^{\alpha\gamma}_{{\mathbf p}_1}\phi^{\beta\gamma}_{{\mathbf p}_2}
\phi^{\delta\mu}_{{\mathbf p}_3}\phi^{\nu\rho}_{{\mathbf p}_4}$
&$U_{22}=U_{\alpha\gamma,\beta\gamma,\delta\mu,\nu\rho}$\\\hline
$\quad 23 \quad$ &
\begin{pspicture}[0.52](2,1)
\rput(0.4,0.25){\dotnode{A}{}}\rput(0.8,0.25){\dotnode{B}{}}
\rput(0.4,0.75){\dotnode{C}{}}\rput(0.8,0.75){\dotnode{D}{}}
\rput(1.2,0.25){\dotnode{M}{}}\rput(1.6,0.25){\dotnode{N}{}}
\rput(1.2,0.75){\dotnode{R}{}}\rput(1.6,0.75){\dotnode{O}{}}
\ncline{A}{B}\ncline{C}{D}\ncline{M}{N}\ncline{R}{O}
\end{pspicture} &
$\displaystyle \bar{u}_{23}\,\sum_{\alpha\beta\gamma\delta\mu\nu\rho\omega}
\phi^{\alpha\beta}_{{\mathbf p}_1}\phi^{\gamma\delta}_{{\mathbf p}_2}
\phi^{\mu\nu}_{{\mathbf p}_3}\phi^{\rho\omega}_{{\mathbf p}_4}$
&$U_{23}=U_{\alpha\beta,\gamma\delta,\mu\nu,\rho\omega}$\\\hline
\end{tabular}
\end{center}
\end{itemize}

\section{The linear relations between the two sets of
vertices}\label{App2}

Decomposing the restricted sums in Eq.\ (\ref{bareG}) and comparison with
(\ref{L}) and (\ref{LI}) provide the linear relations between the upper
and lower case bare couplings. (Although this is straightforward
combinatorics, it becomes quite time consuming
for highly disconnected quartic vertices.) In what follows, these formulae
are displayed by omitting the bars in the notations; in this form they
can be considered as the {\em definition\/} of the exact lower case
vertices. The results for the masses and cubic vertices are taken over
from \cite{rscikk}, and they are shown here for easing to refer them:
\begin{equation}\label{m and w}
\begin{aligned}
 m_1&=\frac{1}{2}(M_1-2M_2+M_3)\\
 m_2&=M_2-M_3\\
 m_3&=\frac{1}{4}M_3;
\end{aligned}\qquad\qquad
\begin{aligned}
w_1&=W_1-3W_5+3W_7-W_8\\
w_2&=\frac{1}{2}W_2-3W_3+\frac{3}{2}W_4+3W_5
+2W_6-6W_7+2W_8\\
w_3&=3W_3-3W_4-6W_5-3W_6+15W_7-6W_8\\
w_4&=\frac{3}{4}W_4-\frac{3}{2}W_7
+\frac{3}{4}W_8\\
w_5&=3W_5-6W_7+3W_8\\
w_6&=W_6-3W_7+2W_8\\
w_7&=\frac{3}{2}W_7-\frac{3}{2}W_8\\
w_8&=\frac{1}{8}W_8.
\end{aligned}
\end{equation}
The set of equations for the quartic case:
\begin{align*}
u_1 & = 3U_1-12U_{16}+12U_{17}+6U_{21}-12U_{22}+3U_{23}  \\
u_2 & = 3U_1+\frac{1}{2}U_2-3U_3+\frac{3}{2}U_4-4U_8+2U_9+12U_{10}+12U_{11}
-24U_{12}+6U_{13}-12U_{14}+12U_{15}\\
\phantom{u_1} & \mathrel{\phantom{=}} -24U_{16}+48U_{17}-24U_{18}-6U_{19}+
24U_{20}+30U_{21}-72U_{22}+18U_{23}\\
u_3 & = -6U_1+3U_3-3U_4-6U_{10}-12U_{11}+12U_{12}+18U_{14}-12U_{15}
+36U_{16}-48U_{17}+12U_{18}\\
\phantom{u_3} & \mathrel{\phantom{=}} +3U_{19}-12U_{20}-39U_{21}+72U_{22}
-18U_{23}\\
u_4 & = \frac{3}{4}U_4-3U_{14}+\frac{3}{2}U_{15}+3U_{21}-3U_{22}
+\frac{3}{4}U_{23}\\
u_5 & = 6U_5-24U_6+12U_7-12U_{11}+12U_{12}-6U_{13}+6U_{14}-6U_{15}
+36U_{16}-84U_{17}+48U_{18}\\
\phantom{u_5} & \mathrel{\phantom{=}} -24U_{20}-36U_{21}+96U_{22}-24U_{23}\\
u_6 & = 12U_6-12U_7-24U_{16}+72U_{17}-24U_{18}+12U_{20}+24U_{21}-84U_{22}
+24U_{23}\\
u_7 & = 2U_7-6U_{17}+6U_{22}-2U_{23}\\
u_8 & = 4U_8-4U_9-12U_{10}-12U_{11}+48U_{12}-12U_{13}+12U_{14}
-24U_{15}+24U_{16}-96U_{17}+48U_{18}\\
\phantom{u_8} & \mathrel{\phantom{=}} +8U_{19}-56U_{20}-48U_{21}+168U_{22}
-48U_{23}\displaybreak\\
u_9 & = U_9-6U_{12}+3U_{15}+6U_{17}+4U_{20}-12U_{22}+4U_{23}\\
u_{10} & = 6U_{10}-12U_{12}-6U_{14}+12U_{15}-12U_{16}+48U_{17}
-12U_{18}-6U_{19}+24U_{20}+30U_{21}-108U_{22}\\
\phantom{u_10} & \mathrel{\phantom{=}}  +36U_{23}\\
u_{11} & = 12U_{11}-12U_{12}-12U_{14}+12U_{15}-24U_{16}+48U_{17}
-12U_{18}+12U_{20}+36U_{21}-84U_{22}+24U_{23}\\
u_{12} & = 6U_{12}-6U_{15}-12U_{17}-6U_{20}+30U_{22}-12U_{23}\\
u_{13} & = -12U_{12}+6U_{13}+6U_{15}+36U_{17}-24U_{18}+24U_{20}
+12U_{21}-72U_{22}+24U_{23}\\
u_{14} & = 3U_{14}-3U_{15}-6U_{21}+9U_{22}-3U_{23}\\
u_{15} & = \frac{3}{4}U_{15}-6U_{22}+\frac{9}{4}U_{23}\\
u_{16} & = 12U_{16}-24U_{17}-12U_{21}+36U_{22}-12U_{23}\\
u_{17} & = 6U_{17}-12U_{22}+6U_{23}\\
u_{18} & = -24U_{17}+12U_{18}-12U_{20}-12U_{21}+60U_{22}-24U_{23}\\
u_{19} & = U_{19}-4U_{20}-3U_{21}+12U_{22}-6U_{23}\\
u_{20} & = 2U_{20}-6U_{22}+4U_{23}\\
u_{21} & = 3U_{21}-6U_{22}+3U_{23}\\
u_{22} & = \frac{3}{2}U_{22}-\frac{3}{2}U_{23}\\
u_{23} & =\frac{1}{16}U_{23}.
\end{align*}

\section{The $G^{(k)}$ coefficients, $k=2,3$ and $4$, in terms of the vertices}
\label{App3}

Eqs.\ (\ref{G^(k+1)}) and (\ref{vertices}) define the $G^{(k)}$'s, which
build up $g(v)$. [See also the paragraph below (\ref{vertices}).] For
$k=2$, it was worked out in Section\ \ref{symmetry}, see (\ref{G^(2)}) and
(\ref{k=0}):
\[
-G^{(2)}=\big[2m_1+n\,m_2\big]+p(n-p)\,4m_3.
\]
We only quote the results for $k=3$ and $k=4$:
\begin{align}\label{W0}
G^{(3)}&=\big[2w_2+n\,w_3+n^2\,w_6\big]+p(n-p)\,\big[(4w_4+2w_5-2w_6)+n\,2w_7\big]
+p^2(n-p)^2\,8w_8\\
\intertext{and}\label{U0}
G^{(4)}&=\big[2u_2+n\,(u_3+u_8)+n^2\,u_{10}+n^3\,u_{19}\big]
+p(n-p)\,\big[(2u_1+4u_4+4u_9-2u_{10}+2u_{11}+2u_{13})\notag\\
\phantom{G^{(4)}} & \mathrel{\phantom{=}} +n\,(2u_{12}+2u_{14}
+u_{16}+u_{18}-3u_{19})+n^2\,(2u_{20}+u_{21})\big]\\
\phantom{G^{(4)}} & \mathrel{\phantom{=}} 
+p^2(n-p)^2\,\big[(8u_{15}+4u_{17}-4u_{20}+24u_{22}+384u_{23})+n\,4u_{22}\big]
+p^3(n-p)^3\,16u_{23}.\notag
\end{align}

From these examples, we can observe the following properties valid also for
generic $k$:
\begin{itemize}
\item $G^{(k)}$ is a double polinomial of $p(n-p)$ and $n$. We must, however,
notice that the vertices themselves depend on $n$. When constructing $g(v)$
by its definition (\ref{g}), we are particularly interested in the limit
$p(n-p)\to 0$ while $n$ fixed, and then taking the spin glass limit $n\to 0$.
\item This limit results the fully replicon vertices $2m_1$, $2w_2$, $2u_2$,
\ldots\ etc. (These are represented by simple two-node graphs connected by
$k$ edges. See the graphical representations in the second column of the
tables in Appendix \ref{App1}.) This justifies, at least in the spin glass
limit, the name replicon generator for $g(v)$.
\end{itemize}

\section{The Legendre-transform method to derive
Eq.\ (\ref{basic})}\label{App4}

The Legendre-transformed free energy $\tilde F(v_{\alpha\beta})$
for the model of Eqs.\ (\ref{bareM})
and (\ref{bareG}), or equivalently (\ref{L}) and (\ref{LI}), is defined
by the usual rules
\begin{equation}\label{Legendre}
-\tilde F=\ln Z-N\sum_{(\alpha\beta)}H_{\alpha\beta}\,v_{\alpha\beta}\quad
\text{and}\quad \frac{\partial(\ln Z)}{\partial H_{\alpha\beta}}=
N v_{\alpha\beta},
\end{equation}
where the partition function $Z=\int\mathcal{D}\phi\, e^{-\mathcal{L}}$
acquires its dependence on the $H_{\alpha\beta}$'s by adding the source term
$-N^{\frac{1}{2}}\sum_{(\alpha\beta)}
H_{\alpha\beta}\,\phi^{\alpha\beta}_{{\mathbf p}=0}$ to the Lagrangian. By
the second equation of (\ref{Legendre}) $\sqrt{N}
\,v_{\alpha\beta}=
\langle \phi^{\alpha\beta}_{{\mathbf p}=0} \rangle$ follows (the average is
taken by the measure $e^{-\mathcal{L}}$), and by the following redefinition
of the fields we can make the 1-point function disappear:
\begin{equation}\label{varphi}
\varphi^{\alpha\beta}_{{\mathbf p}}=\phi^{\alpha\beta}_{{\mathbf p}}-
\sqrt{N}
\,v_{\alpha\beta}\,\,\delta^{\text{Kr}}_{{\mathbf p}=0},
\quad \text{resulting}\quad \langle \varphi^{\alpha\beta}_{{\mathbf p}} \rangle
=0.
\end{equation}
The reader is now reminded that the bare couplings of the original sourceless
--- i.e.\ $H_{\alpha\beta}=0$ --- system are defined likewise,
$\langle \phi^{\alpha\beta}_{{\mathbf p}=0} \rangle=0$ [see the remark in the
paragraph below (\ref{LI})]. Hence we can see that $v_{\alpha\beta}=0$ if
$H_{\alpha\beta}=0$, and it follows from the inverse expression
$\frac{\partial \tilde F}{\partial v_{\alpha\beta}}=NH_{\alpha\beta}$
that stationarity of $\tilde F$ occurs for $v_{\alpha\beta}\equiv 0$.

For computing $\tilde F$, $\phi$ is replaced by $\varphi$,
according to Eq.\ (\ref{varphi}), in the Lagrangian [(\ref{bareM}) and
(\ref{bareG}), supplemented by the source] to give from (\ref{Legendre})
\begin{multline}\label{tilde1}
-\frac{1}{N}\tilde F=\frac{1}{N}\ln Z-\sum_{(\alpha\beta)}H_{\alpha\beta}
\,v_{\alpha\beta}
=\frac{1}{N}\ln \tilde Z\\
+\bar H \,\sum_{(\alpha\beta)}
v_{\alpha\beta}-\frac{1}{2}\sum_{(\alpha\beta),(\gamma\delta)}
\bar{M}_{\alpha\beta,\gamma\delta}\,\,v_{\alpha\beta}v_{\gamma\delta}
+\sum_{k=3}
\,\frac{1}{k!}\,
\sum_{(\alpha_1\beta_1),\dots ,(\alpha_k\beta_k)}
\bar{G}^{(k)}_{\alpha_1\beta_1,\dots ,\alpha_k\beta_k}\,
v_{\alpha_1\beta_1}\ldots\,
v_{\alpha_k\beta_k}.
\end{multline}
The new Lagrangian $\tilde{\mathcal{L}}$ of the shifted system gives
$\ln \tilde Z$, its coupling parameters are listed below up to cubic order:
\begin{equation}\label{tilded couplings}
\begin{aligned}
&\tilde H_{\alpha\beta}=(\bar H+H_{\alpha\beta})-
\sum_{(\gamma\delta)}\bar{M}_{\alpha\beta,\gamma\delta}\,\,v_{\gamma\delta}
+\frac{1}{2}\sum_{(\gamma\delta),(\mu\nu)}
\bar{W}_{\alpha\beta,\gamma\delta,\mu\nu}\,\,v_{\gamma\delta}v_{\mu\nu}
\\&\phantom{\tilde H_{\alpha\beta}}
+\frac{1}{6}\sum_{(\gamma\delta),(\mu\nu),(\rho\omega)}
\bar{U}_{\alpha\beta,\gamma\delta,\mu\nu,\rho\omega}\,\,
v_{\gamma\delta}v_{\mu\nu}v_{\rho\omega}
+\dots, \\[5pt]
&\tilde M_{\alpha\beta,\gamma\delta}=\bar M_{\alpha\beta,\gamma\delta}
-\sum_{(\mu\nu)}\bar{W}_{\alpha\beta,\gamma\delta,\mu\nu}\,\,v_{\mu\nu}
-\frac{1}{2}\sum_{(\mu\nu),(\rho\omega)}
\bar{U}_{\alpha\beta,\gamma\delta,\mu\nu,\rho\omega}\,\,
v_{\mu\nu}v_{\rho\omega}
+\dots, \\[5pt]
&\tilde W_{\alpha\beta,\gamma\delta,\mu\nu}=
\bar W_{\alpha\beta,\gamma\delta,\mu\nu}
+\sum_{(\rho\omega)}\bar{U}_{\alpha\beta,\gamma\delta,\mu\nu,\rho\omega}\,\,
v_{\rho\omega}
+\dots\quad. 
\end{aligned}
\end{equation}
Since $H_{\alpha\beta}$ enters $\tilde{\mathcal{L}}$ solely via $\tilde H$
and the condition in (\ref{varphi}) guarantees the lack of 1-point insertions,
$\tilde F$ depends on the $v_{\alpha\beta}$'s alone.\footnote{Although the
derivation of $\tilde F$ presented here was for the RS spin glass
field theory, the method is entirely general: instead of $(\alpha\beta)$,
a generic index set for the field components could have been used,
and the symmetry of the model does not effectively matter.} 

As it is well known from the literature (see \cite{Amit,Brezin_et_al}, for
instance, and references therein), $\tilde F$ is the generator function
for the exact vertices, and since
\[
\tilde F(v_{\alpha\beta})=F(q+v_{\alpha\beta})+[q\text{-dependent term}],
\]
[cf.\ with (\ref{modF})], we have by Eq.\ (\ref{vertices}):
\begin{multline}\label{tilde2}
-\frac{1}{N}\tilde F=\frac{1}{N}\ln Z(H_{\alpha\beta}\equiv 0)\\
-\frac{1}{2}\sum_{(\alpha\beta),(\gamma\delta)}
{M}_{\alpha\beta,\gamma\delta}\,\,v_{\alpha\beta}v_{\gamma\delta}
+\sum_{k=3}
\,\frac{1}{k!}\,
\sum_{(\alpha_1\beta_1),\dots ,(\alpha_k\beta_k)}
{G}^{(k)}_{\alpha_1\beta_1,\dots ,\alpha_k\beta_k}\,
v_{\alpha_1\beta_1}\ldots\,
v_{\alpha_k\beta_k}.
\end{multline}
Proceeding further along the same lines with Section\ \ref{Mean field theory},
the one-variable function $\tilde F(v)$ is defined by taking
$v_{\alpha\beta}=v$ for $\alpha\cap\beta=0$, otherwise $v_{\alpha\beta}$
is $0$.
From the two alternative forms for $\tilde F$ in Eqs.\ (\ref{tilde1})
and (\ref{tilde2}), and
using the generic definitions (\ref{g}) and (\ref{G^(k+1)}), our basic
result for $g(v)$, displayed in Eq.\ (\ref{basic}), follows.

\section{Block diagonalizing a matrix with the ``two-packet'' 
replica symmetry breaking}\label{App5}

In this Appendix the greek-roman notation of Bray and Moore \cite{BrMo79}
is used to ease the classification of matrix elements when replica
symmetry is broken by dividing the $n$ replicas into two groups:
the $p$ replicas in the first group are denoted by $\alpha$, $\beta$,
\dots, while the other $n-p$ by $a$, $b$, \ldots.
For the 15 different matrix elements the following concise notations will be
used:
\begin{itemize}
\item First class (diagonal elements):
\[
M_1^{(1)}=M_{\alpha a,\alpha a}\qquad\qquad\qquad 
\begin{aligned}
M_1^{(2)}&=M_{\alpha\beta,\alpha\beta}\\
{M'_1}^{(2)}&=M_{ab,ab}
\end{aligned}
\]
\item Second class (one common replica):
\[
\begin{aligned}
M_2^{(1)}&=M_{a\gamma,b\gamma}\\
{M'_2}^{(1)}&=M_{\alpha c,\beta c}
\end{aligned}\qquad\qquad
\begin{aligned}
M_2^{(2)}&=M_{\alpha\gamma,\beta\gamma}\\
{M'_2}^{(2)}&=M_{ac,bc}
\end{aligned}\qquad\qquad
\begin{aligned}
M_2^{(3)}&=M_{\alpha\gamma,b\gamma}\\
{M'_2}^{(3)}&=M_{\alpha c,bc}
\end{aligned}
\]
\item Third class (all replicas are different):
\[
M_3^{(1)}=M_{\alpha\beta,ab}\qquad\qquad
\begin{aligned}
M_3^{(2)}&=M_{\alpha\beta,\gamma\delta}\\
{M'_3}^{(2)}&=M_{ab,cd}
\end{aligned}\qquad\qquad
\begin{aligned}
M_3^{(3)}&=M_{\alpha\beta,\gamma d}\\
{M'_3}^{(3)}&=M_{\alpha b,cd}
\end{aligned}\qquad\qquad
M_3^{(4)}=M_{\alpha b,\gamma d}
\]
\end{itemize}

The procedure of Ref.\ \cite{block_diag} can be applied to find the
block diagonalized matrix $\hat M$ by a similarity transformation
(see also \cite{rscikk}). The
basis vectors%
\footnote{Any elements of the basis vectors not indicated explicitly are zero.}
of the invariant subspaces, matrix blocks and multiplicities
are listed below:
\begin{enumerate}
\item The 3-dimensional longitudinal subspace:
\[
\phi_{(\text{L}0)}^{\alpha a}=1\qquad \qquad\phi_{(\text{L}1)}^{\alpha\beta}=1
\qquad\qquad \phi_{(\text{L}1')}^{ab}=1
\]
\vspace{8pt}
\begin{align*}
{\hat M}_{(\text{L}0),(\text{L}0)}&=M_1^{(1)}+(n-p-1)M_2^{(1)}+
(p-1){M'_2}^{(1)}+(p-1)(n-p-1)M_3^{(4)}\\[3pt]
{\hat M}_{(\text{L}0),(\text{L}1)}&=\frac{1}{2}(p-1)\,\big[(p-2)M_3^{(3)}+
2M_2^{(3)}\big]\\[3pt]
{\hat M}_{(\text{L}0),(\text{L}1')}&=\frac{1}{2}(n-p-1)\,\big[(n-p-2){M'_3}^{(3)}+
2{M'_2}^{(3)}\big]\\[12pt]
{\hat M}_{(\text{L}1),(\text{L}0)}&=(n-p)\,\big[2M_2^{(3)}+(p-2)M_3^{(3)}\big]
\\[3pt]
{\hat M}_{(\text{L}1),(\text{L}1)}&=M_1^{(2)}+2(p-2)M_2^{(2)}+
\frac{1}{2}(p-2)(p-3)M_3^{(2)}\\[3pt]
{\hat M}_{(\text{L}1),(\text{L}1')}&=\frac{1}{2}(n-p)(n-p-1)M_3^{(1)}\\[12pt]
{\hat M}_{(\text{L}1'),(\text{L}0)}&=p\,\big[2{M'_2}^{(3)}+(n-p-2){M'_3}^{(3)}
\big]\\[3pt]
{\hat M}_{(\text{L}1'),(\text{L}1)}&=\frac{1}{2}p(p-1)M_3^{(1)}\\[3pt]
{\hat M}_{(\text{L}1'),(\text{L}1')}&={M'_1}^{(2)}+2(n-p-2){M'_2}^{(2)}+
\frac{1}{2}(n-p-2)(n-p-3){M'_3}^{(2)}
\end{align*}
\vspace{8pt}
\[
\text{Multiplicity:}\quad 3
\]
\item The 2-dimensional anomalous (A) subspaces:

\begin{align*}
\phi_{(\mu 0)}^{\alpha a}&=
\begin{cases}
1 & \text{if $\alpha\not=\mu$}\\
-(p-1)& \text{if $\alpha=\mu$}
\end{cases}\qquad&\qquad
\phi_{(\mu 1)}^{\alpha\beta}&=
\begin{cases}
1 & \text{if $\alpha$, $\beta\not=\mu$}\\
-\frac{1}{2}(p-2)& \text{if $\alpha=\mu$ or $\beta=\mu$}
\end{cases}\\[14pt]
\phi_{(m 0)}^{\alpha a}&=
\begin{cases}
1 & \text{if $a\not=m$}\\
-(n-p-1)& \text{if $a=m$}
\end{cases}\qquad&\qquad
\phi_{(m 1)}^{ab}&=
\begin{cases}
1 & \text{if $a$, $b\not=m$}\\
-\frac{1}{2}(n-p-2)& \text{if $a=m$ or $b=m$}
\end{cases}
\end{align*}
\[
\begin{gathered}
\mu\text{-block:}\\[110pt]
m\text{-block:}
\end{gathered}\qquad\qquad
\begin{aligned}
{\hat M}_{(\mu 0),(\mu 0)}&=M_1^{(1)}+(n-p-1)M_2^{(1)}-{M'_2}^{(1)}-
(n-p-1)M_3^{(4)}\\[3pt]
{\hat M}_{(\mu 0),(\mu 1)}&=\frac{1}{2}(p-2)\,\big(M_2^{(3)}-M_3^{(3)}\big)\\[10pt]
{\hat M}_{(\mu 1),(\mu 0)}&=2(n-p)\,\big(M_2^{(3)}-M_3^{(3)}\big)\\[3pt]
{\hat M}_{(\mu 1),(\mu 1)}&=M_1^{(2)}+(p-4)M_2^{(2)}-(p-3)M_3^{(2)}\\[24pt]
{\hat M}_{(m 0),(m 0)}&=M_1^{(1)}+(p-1){M'_2}^{(1)}-{M_2}^{(1)}-
(p-1)M_3^{(4)}\\[3pt]
{\hat M}_{(m 0),(m 1)}&=\frac{1}{2}(n-p-2)\,\big({M'_2}^{(3)}-{M'_3}^{(3)}\big)\\[10pt]
{\hat M}_{(m 1),(m 0)}&=2p\,\big({M'_2}^{(3)}-{M'_3}^{(3)}\big)\\[3pt]
{\hat M}_{(m 1),(m 1)}&={M'_1}^{(2)}+(n-p-4){M'_2}^{(2)}-(n-p-3){M'_3}^{(2)}
\end{aligned}
\]
\vspace{8pt}
\[
\text{Multiplicity:}\quad 2\,\big[(p-1)+(n-p-1)\big]=2(n-2)
\]

\item The 1-dimensional R0 replicon subspaces:

\[
\phi_{(\mu m)}^{\alpha a}=
\begin{cases}
1 & \text{if $\alpha\not=\mu$ and $a\not=m$}\\
-(p-1) & \text{if $\alpha=\mu$ and $a\not=m$}\\
-(n-p-1) & \text{if $\alpha\not=\mu$ and $a=m$}\\
(p-1)(n-p-1) & \text{if $\alpha=\mu$ and $a=m$}
\end{cases}
\]
\vspace{8pt}
\[
{\hat M}_{(\mu m),(\mu m)}=M_1^{(1)}-M_2^{(1)}-{M'_2}^{(1)}+M_3^{(4)}
\]
\vspace{8pt}
\[
\text{Multiplicity:}\quad (p-1)(n-p-1)=p(n-p)-(n-1)
\]

\item The 1-dimensional R1 replicon subspaces:

\begin{itemize}
\item $(\mu\nu)$ eigenvector:
\[
\phi_{(\mu\nu)}^{\alpha\beta}=
\begin{cases}
1 & \text{if $\alpha$, $\beta\not=\mu$, $\nu$}\\
-\frac{1}{2}(p-3) & \text{if one of $\alpha$ and $\beta$ coincides
with one of $\mu$ and $\nu$}\\
\frac{1}{2}(p-2)(p-3) & \text{if $\alpha=\mu$ and $\beta=\nu$,
or $\alpha=\nu$ and $\beta=\mu$}
\end{cases}
\]
\vspace{8pt}
\[
{\hat M}_{(\mu\nu),(\mu\nu)}=M_1^{(2)}-2M_2^{(2)}+M_3^{(2)}
\]
\vspace{8pt}
\[
\text{Multiplicity:}\quad \frac{1}{2}p(p-3)
\]

\vspace{10pt}
\item $(m n)$ eigenvector:
\[
\phi_{(mn)}^{ab}=
\begin{cases}
1 & \text{if $a$, $b\not=m$, $n$}\\
-\frac{1}{2}(n-p-3) & \text{if one of $a$ and $b$ coincides
with one of $m$ and $n$}\\
\frac{1}{2}(n-p-2)(n-p-3) & \text{if $a=m$ and $b=n$,
or $a=n$ and $b=m$}
\end{cases}
\]
\vspace{8pt}
\[
{\hat M}_{(mn),(mn)}={M'_1}^{(2)}-2{M'_2}^{(2)}+{M'_3}^{(2)}
\]
\vspace{8pt}
\[
\text{Multiplicity:}\quad \frac{1}{2}(n-p)(n-p-3)
\]
\end{itemize}
\end{enumerate}

\bibliography{spinglass}

\begin{acknowledgments}
A correspondence with Pierluigi Contucci concerning exact results in the
Edwards-Anderson model is appreciated.
\end{acknowledgments}
\end{document}